\documentclass[aip,jcp,showpacs,nobibnotes,reprint,floatfix,superscriptaddress]{revtex4-1}
\usepackage{graphics,graphicx,amsfonts,amsmath,amsbsy,amssymb,color}
\usepackage{psfrag}  
\usepackage{tabularx}
\usepackage{xmpmulti}
\usepackage{hyperref}
\usepackage{marginnote}
\usepackage{subfigure}
\usepackage{paracol}
\usepackage{xparse}
\usepackage{layouts}
\usepackage{afterpage}
\usepackage{braket}
\usepackage{paralist}
\usepackage{bm}
\usepackage{url}
\usepackage[normalem]{ulem}
\usepackage{color}
\usepackage[colorinlistoftodos]{todonotes}
\usepackage[american]{babel} 

\newcommand{\refeq}[1]{{Eq.~(\ref{#1})}}
\newcommand{\reffig}[1]{{Fig.~\ref{#1}}}
\newcommand{\refsec}[1]{{Sec.~\ref{#1}}}

\newcommand{\refapp}[1]{{Appendix~\ref{#1}}}

\hypersetup{hyperindex,breaklinks,colorlinks=true,linkcolor=blue,citecolor=blue,urlcolor=blue,filecolor=blue}

\begin{document}

%
% revtex 4-1 
\title{The sign problem in density matrix quantum Monte Carlo}
\author{Hayley~R.~Petras}
\altaffiliation{These authors contributed equally to this paper}
\address{Department of Chemistry, University of Iowa}
\author{William~Z.~Van Benschoten}
\altaffiliation{These authors contributed equally to this paper}
\address{Department of Chemistry, University of Iowa}
\author{Sai~Kumar~Ramadugu}
\address{Department of Chemistry, University of Iowa}
\author{James~J.~Shepherd}
\email{james-shepherd@uiowa.edu}
\address{Department of Chemistry, University of Iowa}
%
% end revtex
%
%
\begin{abstract}

Density matrix quantum Monte Carlo (DMQMC) is a recently-developed method for stochastically sampling the $N$-particle thermal density matrix to obtain exact-on-average energies for model and \emph{ab initio} systems. We report a systematic numerical study of the sign problem in DMQMC based on simulations of atomic and molecular systems. In DMQMC, the density matrix is written in an outer product basis of Slater determinants and has a size of space which is the square of the number of Slater determinants. In principle this means DMQMC needs to sample a space which scales in the system size, $N$, as $\mathcal{O}[(\exp(N))^2]$. In practice, there is a system-dependent critical walker population ($N_c$) which must be exceeded in order to remove the sign problem, and this imposes limitations by way of storage and computer time. 
We establish that $N_c$ for DMQMC is the square of $N_c$ for FCIQMC. By contrast, the minimum $N_c$ in the interaction picture modification of DMQMC (IP-DMQMC) only is directly proportionate to the $N_c$ for FCIQMC. 
We find that this comes from the asymmetric propagation of IP-DMQMC compared to the symmetric propagation of canonical DMQMC.
An asymmetric mode of propagation is prohibitively expensive for DMQMC because it has a much greater stochastic error.
Finally, we find that the equivalence between IP-DMQMC and FCIQMC seems to extend to the initiator approximation, which is often required to study larger basis sets and other systems. This suggests IP-DMQMC offers a way to ameliorate the cost of moving between a Slater determinant space and an outer product basis.
\end{abstract}
\date{\today}
\maketitle

\section{Introduction}

In a recent study, we showed that the density matrix quantum Monte Carlo (DMQMC) method could be applied to molecular systems, extending it beyond original applications to model systems in condensed matter physics.\cite{petras_using_2020} The use of finite temperature electronic structure methods are becoming increasingly important in applications such as plasmonic catalysis,\cite{mukherjee_hot_2013, zhou_aluminum_2016} the study of planetary interiors,\cite{mazzola_phase_2018} and solid-state materials\cite{gull_superconductivity_2013}, where the temperature dependence is key in obtaining physical and chemical properties, such as phase diagrams and excitation energies. The inclusion of temperature in quantum chemistry methods is difficult because at finite temperatures, more than one state is often occupied, increasing the difficulty of solving the Schrodinger equation. 
DMQMC joins a growing set of methods including other quantum Monte Carlo methods,\cite{liu_ab_2018,liu_unveiling_2020, ceperley_fermion_1991, ceperley_path-integral_1992,dornheim_ab_2018, leblanc_magnetic_2019} many body theories\cite{sanyal_thermal_1992,li_multilevel_2010,he_finite-temperature_2014} and others in attempting to solve the finite temperature problem that has attracted recent attention amongst quantum chemists. \cite{rusakov_self-consistent_2016,doran_monte_2019, neuhauser_stochastic_2017, hirschmeier_mechanisms_2015, hirshberg_path_2020} Many of these methods, like DMQMC, continue to undergo development.\cite{yilmaz_restricted_2020, dornheim_permutation_2019, harsha_thermofield_2019-1, harsha_thermofield_2019, shushkov_real-time_2019, white_time-dependent_2018, white_finite-temperature_2020, hummel_finite_2018, roggero_quantum_2013} 

Widespread adoption of all methods in the FCIQMC family, including DMQMC, is hindered, in part, due to the sign problem.
In FCIQMC-based methods, coefficients in the wavefunction (or density matrix, in DMQMC) are sampled by a distribution of walkers. 
The original FCIQMC paper found that simulations that exceeded a critical walker population were able to successfully resolve the signs of the wavefunction and generate an energy estimate that was exact-on-average; it was not possible to find accurate estimates from populations lower than the plateau.\cite{booth_fermion_2009}
In FCIQMC, walkers arriving at the same site can be exactly annihilated due to having the discrete basis set; this contrasts a continuous real-space basis where the same approach can be much more difficult. 
While the wavefunction is still being sampled exactly on average, the signal-to-noise ratio is extremely low and prevents exact estimates from being extracted.
A simulation with a growing walker population will have its growth briefly stall out, forming a plateau in the total walker population ($N_w$) as a function of the simulation iteration, known as the ``annihilation plateau", as the simulation establishes the sign of critical elements of the wavefunction.
When the population has grown above the plateau, the sign problem is resolved and exact energies can be straight-forwardly collected.

The sign problem in FCIQMC was discussed in depth in the early developmental papers\cite{booth_fermion_2009, booth_breaking_2011} for the method before subsequently being systematically studied by Spencer \emph{et al.},\cite{spencer_sign_2012} whose work we refer to throughout. 
This work established the origin of the sign problem as an unphysical Hamiltonian (whose solution does not have a sign problem) and that is unavoidably encountered in undersampled dynamics.
There are also attempts to leverage this understanding directly using a fixed-node or trial wavefunction approach.\cite{kolodrubetz_partial_2012,roggero_quantum_2013}
The development of the initiator approach in FCIQMC removed the annihilation plateau at a cost of introducing a small error in the energy (removed by increasing the number of walkers in the simulation).\cite{cleland_communications:_2010}
The motivation for and derivation of this approximation was related to the alleviation of the sign problem and allowed for a much broader scope of applications.
The development of the initiator approximation in DMQMC achieved a similar outcome allowing for our previous work on the uniform electron gas and \emph{ab initio} molecular systems.\cite{malone_accurate_2016,petras_using_2020}
Subsequently, there were also a wide variety of FCIQMC or FCIQMC-like methods development which are beyond the scope of this work to review in detail.\cite{deustua_communication:_2018,blunt_hybrid_2019,ghanem_unbiasing_2019,ghanem_adaptive_2020,anderson_four-component_2020,vitale_fciqmc-tailored_2020, anderson_efficient_2020,li_manni_compression_2020, petras_fully_2019, dobrautz_efficient_2019, blunt_preconditioning_2019, luo_combining_2018,blunt_communication:_2018,  li_manni_combining_2016,tubman_deterministic_2016, blunt_semi-stochastic_2015} Large scale implementations of the FCIQMC method and related methods have also been developed and these papers review current challenges and developments for the interested reader.\cite{spencer_hande-qmc_2019, guther_neci_2020}

Here, we conduct a systematic investigation of the sign problem in density matrix quantum Monte Carlo (DMQMC).
We find that the annihilation plateau comes from the same unphysical Hamiltonian as in FCIQMC. We measure these critical walker populations for a test set from the FCIQMC literature and find that DMQMC plateau heights are proportional to the square of the FCIQMC plateau height. However, we also show that by moving to the interaction picture (IP-DMQMC) the plateau heights scale linearly with the FCIQMC plateau heights. 
Despite being able to control the sign problem, IP-DMQMC can have an issue with the trace population as there is no global estimator for the energy in DMQMC unlike in FCIQMC. To address the collapse of the trace population that occurs even when the sign problem is overcome, we examine the initiator adaptation, showing that it has similar performance as the initiator adaptation in ground-state calculations using FCIQMC.

We find that the reason that IP-DMQMC has this plateau height reduction is that the propagation is asymmetric. 
Comparing asymmetric DMQMC to IP-DMQMC, we find that the critical populations are \emph{the same} when a shift is used in DMQMC. 
While asymmetric DMQMC does appear to have cost savings in the required population compared to symmetric DMQMC, this is offset by the need to sample over the rows (or columns), or, equivalently, $\beta$-loops. 
We believe this shows IP-DMQMC is as effective at solving for finite-temperature energies as FCIQMC is at solving zero-temperature energies.
We see this work as complementary to our previous and future studies which develop and apply DMQMC as well as the related work of Rubenstein \emph{et al.} discussing the sign problem for finite-temperature auxiliary field quantum Monte Carlo.\cite{shen_finite_2020, liu_unveiling_2020, liu_ab_2018} 

\section{Methods} \label{methods}
In this section, we provide a summary of the methods used here. We begin with the three methods primarily used in this work: DMQMC, interaction picture DMQMC and FCIQMC. We then describe the initiator adaptation. 
We note now that Hartree atomic units are used throughout.

\subsection{Density matrix quantum Monte Carlo}\label{dmqmc-methods}
We begin with the original formulation of DMQMC.\cite{blunt_density-matrix_2014}  Starting with the unnormalized thermal density matrix
\begin{equation}
\hat{\rho} = e^{-\beta \hat{H}}
\end{equation}
where $\hat{H}$ is the Hamiltonian operator and $\beta = ({k_B T})^{-1}$, we can show that the density matrix satisfies the symmeterized Bloch equation 
\begin{equation}\label{eq:2}
    \frac{d\hat{\rho}}{d\beta} = - \frac{1}{2} (\hat{H}\hat{\rho}+ \hat{\rho}\hat{H}) .
\end{equation}
by differentiating $\hat{\rho} (\beta)$ with respect to $\beta$. 
 A Euler update scheme, or finite difference approach, with a finite time step, $\Delta \beta$, can then be used to find the density matrix at any $\beta$, following 
 \begin{equation}\label{finite-diff}
\hat{\rho} (\beta + \Delta \beta) = \hat{\rho}(\beta) -\frac{\Delta\beta}{2}(\hat{H}\hat{\rho}(\beta) + \hat{\rho}(\beta) \hat{H}) + O(\Delta\beta ^2) .
 \end{equation}

We then rewrite \refeq{finite-diff} in a basis of outer products of Slater determinants to obtain a matrix form that can be solved stochastically, by evolving a population of particles through the inverse temperature regime. The result is 
\begin{equation}\label{matrix}
    \rho_{\mathbf{ij}} (\beta + \Delta\beta) = \rho_{\mathbf{ij}}(\beta) + \frac{\Delta \beta}{2} \sum_{\mathbf{k}} (T_{\mathbf{ik}} \rho_{\mathbf{kj}} + \rho_{\mathbf{ik}}T_{\mathbf{kj}}) 
\end{equation}  
where $T_{\mathbf{ij}} = -(H_{\mathbf{ij}}-S \delta_{\mathbf{ij}})$ is the update matrix, and $S$ is a variable shift for population control of the particles in the simulation, explained later in this section. 

The matrix elements $\rho_{\mathbf{ij}} = \langle D_i | \hat{\rho} | D_j \rangle$ are represented by particles in the simulation, where $|D_i \rangle$ are Slater determinants in the defined finite basis set. The $i$ and $j$ indices begin at $i = 0$ and $j =0$. A population of particles, $N_w$ is then used to sample elements of the density matrix by evolving with respect to $\beta$, according to \refeq{matrix}. Integer weights were used in the original FCIQMC algorithm (which DMQMC is based off of) and because we want to make comparison with previous results, this is what we use here. 

The simulation starts at $\beta=0$, where the density matrix is the identity matrix.  The simulation is then propagated to the desired value of $\beta$.  At each step, the population is updated following rules for spawning and death of particles, summarized below, while particles of opposite signs on each matrix element are annihilated; these steps are closely analogous to FCIQMC.\cite{booth_fermion_2009} 

There are three rules for evolving particles that can described as follows: 
\begin{itemize}
\item Spawning: occurs from one matrix element ($\rho_{\mathbf{ik}}$) to another ($\rho_{\mathbf{ij}}$), along both the rows and columns.   
\item Cloning and death: occur on single matrix elements only, and are designed to increase and decrease the population respectively.
\item Annihilation: particles of opposite signs on single matrix elements are removed from the simulation.
\end{itemize}

Spawning will occur with the probability $p_s(\mathbf{ik} \to \mathbf{ij}) = \frac{\Delta \beta |T_{\mathbf{kj}}|}{2}$ and the sign will correspond to sign($\rho_{\mathbf{ij}}$) = sign$(\rho_{\mathbf{ik}}) \times $sign$(T_{\mathbf{kj}})$.  The same equations will hold for spawning from $\rho_{\mathbf{kj}}$ to $\rho_{\mathbf{ij}}$. 

The sign of the newly spawned particle is important because as can be seen, the sign of the new particle depends on both the sign of the matrix element where the particle spawned \emph{from} and the update matrix, $T$,  connecting the two elements. Because of this, the signs of newly spawned particles will not always be sign-coherent, resulting in the manifestation of the sign problem. In order to resolve the signs of the particles on the matrix elements, a system dependent number of particles is required. This will be explored further throughout this work.

Cloning and death occur with a probability given by $p_d(\mathbf{ij}) = \frac{\Delta \beta}{2} | T_{\mathbf{ii}} + T_{\mathbf{jj}}|$. The population increases if sign$(T_{\mathbf{ii}} + T_{\mathbf{jj}}) \times $sign$(\rho_{\mathbf{ij}}) > 0$, and decreases otherwise.   Annihilation also occurs on single matrix elements and is used to control the sign problem and particle growth within the simulation, and has been show to be key in overcoming the sign problem.\cite{booth_fermion_2009, spencer_sign_2012}

A population control must be used, so we introduce a variable shift parameter, that is controlled by 
\begin{equation}
    S(\beta + A\Delta\beta) = S(\beta) - \frac{\xi}{A\Delta\beta} ln \Big( \frac{N_w (\beta +A\Delta\beta)}{N_w (\beta)}\Big)
\end{equation}
The shift update is dependent on $N_w (\beta)$, the total number of walkers at $\beta$,  $A$, the number of $\Delta \beta$ steps between shift updates, and $\xi$, a shift damping parameter. 

The steps outlined above are repeated until the desired inverse temperature is reached. To obtain estimates of thermodynamic quantities, one averages over many independent simulations, termed ``$\beta$ loops". 
Then, to find the energies the following expression is used, $\langle \hat{H}\rangle={Tr(\hat{\rho}\hat{H})}/{Tr(\hat{\rho})}$, where the numerator and denominator of this equation are sampled separately over the course of the propagation through $\beta$, and averaged over the desired amount of $\beta$ loops. In this work, we solely use the projected estimator and not the shift estimator (because the shift estimator does not converge to the finite temperature energy in DMQMC\cite{blunt_density-matrix_2014}).

\subsection{Interaction Picture DMQMC (IP-DMQMC)} \label{ip-methods}

The interaction picture variant of DMQMC (IP-DMQMC throughout) was developed to overcome two sampling issues present in the original DMQMC method: the initial density matrix rarely contains the important determinants and the distribution of weight fluctuates rapidly as a function of $\beta$.\cite{malone_interaction_2015} 
Replacing the density matrix with an auxiliary matrix, $\hat {f}$, means that the simulation can be started at a non-interacting density matrix, $e^{-\beta \hat{H}^0}$, rather than the identity, providing a good first approximation to the fully interacting density matrix for weakly-correlated systems. The auxiliary matrix can be written: 
\begin{equation}
\hat{f}(\tau) = e^{-(\beta-\tau)\hat{H}^0} e^{-\tau \hat{H}}
\end{equation}
where $\hat{H} = \hat{H}^0 + \hat{V}$ and $\hat{H}^0$ is a mean-field Hamiltonian. %
In this work, we use the Hartree-Fock Hamiltonian for $\hat{H^0}$, though it is possible to use a more general mean-field Hamiltonian. 
In practice, $\hat{H^0}$ only has diagonal matrix elements in a Slater determinant basis, and $e^{-(\beta-\tau)\hat{H}^0}$ only has diagonal matrix elements at any temperature.
It is important to note that this matrix evolves from $e^{-\beta \hat{H}^0}$ at $\tau = 0$ to $e^{-\beta \hat{H}} = \hat{\rho}(\beta)$ at $\tau = \beta$, which means IP-DMQMC only samples the correct distribution at $\tau = \beta$, so separate simulations are required for each $\beta$ value. 

We can differentiate the matrix $\hat{f}$ with respect to $\tau$ to find:
\begin{equation}
\frac{d\hat{f}}{d\tau} = \hat{H}^0 \hat{f} -\hat{f} \hat{H}.
\end{equation}
This equation can be simulated using the rules above, with one change: the cloning/death probability in the second rule changes to $p_d(\mathbf{ij}) = \Delta \tau |H^0_{\mathbf{ii}} - H_{\mathbf{jj}}|$, because $\hat{H^0}$ is diagonal in the chosen basis.
The condition on increasing or decreasing the population is then based on the sign of $(H^0_{\mathbf{ii}} - H_{\mathbf{jj}})\times $sign$(\rho_{\mathbf{ij}})$, with population increasing if this expression is greater than 0. In this work, IP-DMQMC uses the asymmetric spawning mode as described in \refsec{methods-spawning} below, meaning that spawning is restricted to occur only along rows.

IP-DMQMC is the same as DMQMC, in that many simulations need to be averaged to obtain estimates for observables.  
When introduced, it was said that one major benefit of this variant is that as long as $H_{\mathbf{ii}}^0 > H_{\mathbf{jj}} $, there is little to no death along the diagonal; this overcomes one problem with large systems in DMQMC, where the distribution along the diagonal approaches zero with $\beta$.\cite{malone_interaction_2015} 
When $H^0$ is based on Hartree--Fock, $H^0_{\mathbf{ii}}=H_{\mathbf{ii}}$, the initial condition must also be changed and this is described in detail in the original paper.\cite{malone_interaction_2015}
The grand canonical density matrix corresponding to $\hat{H^0}$ is used to obtain the desired distribution according to $e^{-\beta \hat{H}^0}$.

\subsection{Symmetric versus asymmetric spawning}\label{methods-spawning}%

For this study in particular, it is important to distinguish between symmetric and asymmetric modes of spawning. 
In DMQMC as canonically formulated (\refeq{eq:2}) 
spawning is allowed both on rows and columns because the propagator is symmetric. When we refer to DMQMC in this manuscript, we generally mean this canonical formulation unless otherwise specified.
However, it is also possible to have asymmetric DMQMC, with a propagator: 
\begin{equation}
    \frac{d\hat{\rho}}{d\beta} = - \hat{\rho}\hat{H} .
\end{equation}
where the spawning is restricted on rows (or equivalently on columns). %
The propagator in IP-DMQMC is canonically asymmetric, with the same spawning restriction as asymmetric DMQMC. While a symmetric propagator exists for the uniform electron gas,\cite{malone_quantum_2017} 
it does not for molecular systems and is complicated to develop and test.

\subsection{Full configuration interaction quantum Monte Carlo}%

Next, we describe the FCIQMC method~\cite{booth_fermion_2009} in brief as it will be used for comparison throughout this work. 

FCIQMC begins with the imaginary time Schr\"{o}dinger equation
\begin{equation}\label{imag-time}
    \frac{d | \Psi _0 \rangle}{d \tau} = - \hat{H} | \Psi_0 \rangle
\end{equation}
where $| \Psi_0 \rangle$ is the ground state wavefunction, $\hat{H}$ is the Hamiltonian operator and $\tau$ represents imaginary time. Here, the wavefunction is represented as a sum over Slater determinants, $| D_i \rangle$,
\begin{equation}\label{sum-slater}
    | \Psi_0 \rangle = \sum_i c_i |D_i \rangle
\end{equation}
where $c_i$ is the coefficient on the $i^{th}$ determinant
and the Hamiltonian is represented as 
\begin{equation}
    H_{ij} = \langle D_i | \hat{H} | D_j \rangle .
\end{equation}

In the same vein as DMQMC, we can obtain a finite difference equation
\begin{equation}
    c_i^{m+1} - c_i^m =\tau (-H_{ii} +S ) c_i^m  - \sum_{j \neq i} \tau H_{ij} c_j^m
\end{equation}
by substituting the sum over Slater determinants (from \refeq{sum-slater}) into \refeq{imag-time}, where $c_i^m$ is the coefficient of the $i^{th}$ determinant at iteration $m$ of the simulation. Note here that the total population of particles, $N_w$, is given by $N_w = \sum_i | c_i |$. To obtain an estimate of the ground state energy, $S$ is varied to keep the particle population constant, and can be averaged to obtain the estimate. 

The rules for evolving particles are those on which the DMQMC algorithm was subsequently based. At each step of the simulation, the particles on each element will undergo spawning, death/cloning and annihilation, as they do in DMQMC. Particles spawn from site $i$ with weight $c_i$ to connected sites, $j$, where $i \neq j$, where the probability is uniform in $j$. In the death/cloning step, particles on site $i$ increase or decrease their population according to $|S - H_{ii}|\tau$. Particles on site $i$ with opposite signs are removed from the simulation. 

The particle population is  evolved through imaginary time following the rules above, through a system dependent number of iterations. After the wavefunction emerges, the correlation energy is found by averaging over the iterations in the simulation, in a similar fashion to how $\beta$ loops are averaged in DMQMC.

\subsection{Initiator Adaptation}\label{init-methods}

The sparsity of the thermal density matrix (or the wavefunction coefficient matrix in FCIQMC) can be utilized through use of the initiator approximation variation of both methods, here represented as i-DMQMC\cite{malone_accurate_2016} and i-FCIQMC\cite{cleland_communications:_2010,cleland_study_2011}.
The initiator approximation works by setting a threshold ``$n_{add}$" value, where spawning to unoccupied matrix elements is limited to occur only from matrix elements with particle populations larger than $n_{add}$, called ``initiator determinants" (or from co-incident spawns of particles of the same sign from two non-initiator sites).  
This approximation limits the number of density matrix elements (or vector elements in FCIQMC) that need to be sampled over the course of the simulation. Increasing the total number of particles, $N_w$, can reduce the magnitude of the approximation.  Both of the original algorithms are obtained as $N_w \to \infty$. 
The initiator adaptation can be used with or without the interaction picture.

\subsection{Kernel Density Estimation}
The plateau height in this work is defined as the population that occurs with the highest frequency in the simulation, and we call this population the critical population ($N_c$).
The Scott kernel density estimation (KDE) method\cite{scott_multivariate_2014} is used in this work to assign critical populations through a systematic and reproducible protocol. This is a continuous adaptation from prior work.\cite{shepherd_sign_2014}
The KDE method works by calculating the probability that a certain walker population is present in the DMQMC simulation through the use of a KDE kernel, $K$. If we let $f(x)$ be a continuous function representing the population dynamics in one trajectory, we can use the kernel density estimator
\begin{equation}
    \hat{f_h}(x) =  \frac{1}{nh} K \big{(} \frac{x-x_i}{h} \big{)}
\end{equation}
where $h$ is a smoothing parameter and $n$ is the number of data points to find the KDE kernel. The KDE kernel itself gives a probability distribution of the number of walkers as a function of the number of walkers. The maximum value of the kernel will correspond to the critical walker population. 

Simulations to measure the plateau height are performed with a single $\beta$ loop, and the output files are analyzed using the Python scripts provided in the HANDE software package,\cite{spencer_hande-qmc_2019} producing one analysis file per output file. 
The plateau assignments are performed on the data sets with the total walker population ($N_w$) on a logarithmic axis. 
For the plateau heights in DMQMC, there were some cases where the simulations entered variable shift before the annihilation plateau occurred, or the total population collapsed to zero and did not recover. If either of these situations occurred in the simulation, it was not used when measuring plateau heights. 

The maximum value of the KDE kernel is assigned as the critical population, and these are collected in a separate file. Graphs of the KDE kernel and the total walker population are produced, and checked visually to ensure the critical population was assigned correctly. Once all plateaus have been validated by visual inspection, the critical populations are averaged and the standard error calculated. 

We note here that the FCIQMC critical walker populations used throughout this work are from Ref.~\onlinecite{booth_fermion_2009}, and were not recalculated for this work. 

\section{Results and Discussion}\label{results}

Calculations were performed on a variety of linear hydrogen chains, and other small atoms and molecules  using the HANDE-QMC package, version 1.4 and 1.5.\cite{spencer_hande-qmc_2019} All simulations in this work were performed using a timestep of $0.001$ and a shift damping value of $0.30$.  Integral dump files were generated using MOLPRO,\cite{MOLPRO} in the form of an FCIDUMP.\cite{knowles_determinant_1989} The single particle eigenvalues for the systems are then calculated using an in-house code from the orbitals in the FCIDUMP according to standard equations.\cite{szabo_modern_1996}
These single particle eigenvalues are then added to the FCIDUMP before the core Hamiltonian energy.

The equilibrium H$_n$ chains used in this study had a bond length of 0.945110567 Angstroms, and the stretched H$_n$ chains had a bond length of 1.270025398 Angstroms. These correspond to 1.786 a.u. and 2.4 a.u. respectively, which come from a previous study using auxiliary field quantum Monte Carlo.\cite{liu_ab_2018}
The H$_2$O system used a O-H bond length of 0.975512 Angstroms and an H-O-H angle of 110.565 degrees. 
The CH$_4$ system used a C-H bond length of 1.087728 Angstroms, and a H-C-H bond angle of 109.47122 degrees.  The bond lengths for the diatomic systems are as follows: HF, 0.91622 Angstroms; NaH, 1.885977 Angstroms; C$_2$, 1.27273 Angstroms; N$_2$, 2.068 a.u. and stretched N$_2$, 4.2 a.u. These come from a previous study using FCIQMC.\cite{booth_fermion_2009}

The critical walker populations, or ``plateau heights", were measured by the Scott KDE method\cite{scott_multivariate_2014} using NumPy\cite{harris_2020_array} in Python3. 
The DMQMC calculations to measure critical populations for the H$_6$ systems were performed with initial populations of $5 \times 10^2$ and  target populations of $5 \times 10^6$, and for H$_8$, the simulations used initial populations of $5 \times 10^4$ and target populations of $5 \times 10^8$. These simulations were propagated to $\beta$ =25. 
The IP-DMQMC and FCIQMC simulations for measuring the critical populations were performed with an initial population of $1$ and a target population of $5 \times 10^8$. 
All calculations used the integer walker algorithm in all methods to maintain comparability with the plateaus reported in the first FCIQMC paper.\cite{booth_fermion_2009}.  When one walker is used, this means we are sampling exactly one row per $\beta$-loop (for asymmetric methods). %

The following results are now arranged as follows: in \refsec{intro-plateaus}, we begin by confirming the presence of the annihilation plateau and compare the critical walker population in DMQMC to FCIQMC for stretched H$_6$, which essentially reproduces known results from Blunt et al.\cite{blunt_density-matrix_2014}. In \refsec{unphysical-ham}, we then explore the connection to the unphysical Hamiltonian related to FCIQMC.\cite{spencer_sign_2012}
Next, we generalize our finding from \refsec{intro-plateaus} to a wide range of atomic and molecular systems in \refsec{dmqmc-plateau}, and also explore the interaction picture variant of DMQMC. We then discuss similarities and differences between DMQMC and FCIQMC in \refsec{ip-same-as-fciqmc}, and energy convergence in \refsec{Nrows}.
Finally, we compare the initiator adaptions to IP-DMQMC and FCIQMC in \refsec{initiator}. 

Throughout the manuscript, DMQMC refers to symmetric DMQMC. 
This is the only type of DMQMC mentioned in \refsec{intro-plateaus}, \refsec{unphysical-ham}, and \refsec{dmqmc-plateau}.
In \refsec{ip-same-as-fciqmc}, asymmetric DMQMC is introduced and discussed and we continue to use it throughout the paper. 
IP-DMQMC uses asymmetric propagation throughout.
In section headings and the captions of figures, information about whether DMQMC is being propagated in a symmetric or an asymmetric fashion is repeated for emphasis and clarity.

\subsection{An example of a symmetric DMQMC annihilation plateau}\label{intro-plateaus}

\begin{figure}
\begin{center}
   \subfigure[\mbox{}]{ \includegraphics[width=0.45\textwidth,height=\textheight,keepaspectratio]{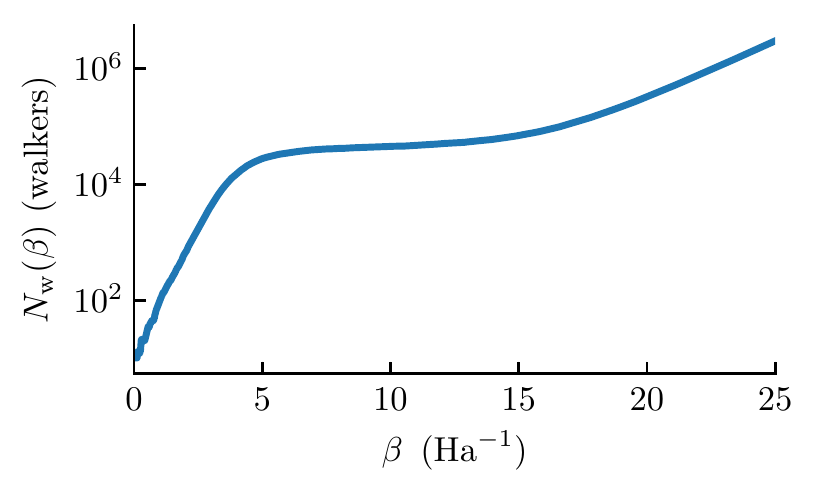} \label{fig1a}}
   \subfigure[\mbox{}]{ \includegraphics[width=0.45\textwidth,height=\textheight,keepaspectratio]{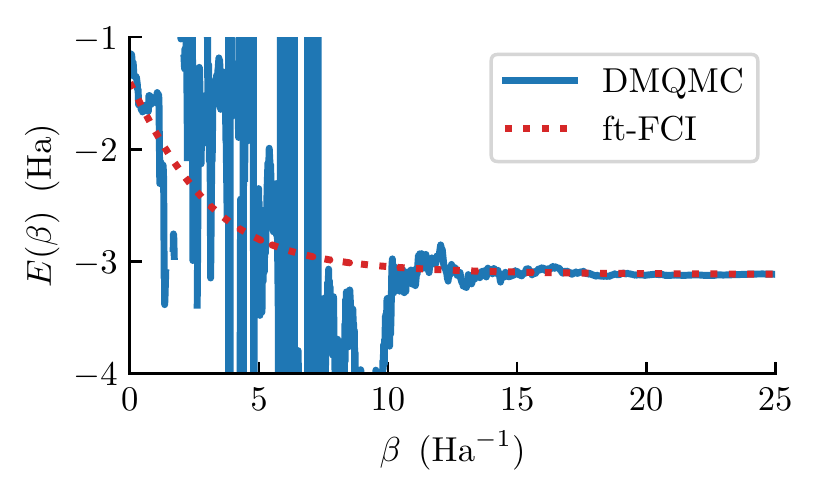}\label{fig1b} }
    \caption{ For stretched H$_6$/STO-3G (a) the total walker population, $N_w(\beta)$  and (b) energy, $E(\beta)$  from a single $\beta$ loop, propagated to $\beta$ = 25. The simulation was started at $\beta=0$, and used a shift of 0.343 to ensure the plateau was exited by $\beta=25$. In (b), the exact diagonalization (ft-FCI) is shown as a red dotted line. These results agree with prior observations.\cite{blunt_density-matrix_2014} In this figure, DMQMC is symmetrically propagated. }
    \label{fig1}
    \end{center}
\end{figure}

We first begin by describing and then reproducing the original finding of the DMQMC annihilation plateau, where we offer an example of an \emph{ab initio} system. 
This section is intended to introduce readers to features of an annihilation plateau.
The first paper on DMQMC\cite{blunt_density-matrix_2014} described the sign problem in this method as similar to that of FCIQMC, due to the close similarities between the population dynamics within the method. 
Of particular interest is the annihilation step, which is identical between the two methods, and is found to be key in overcoming the sign problem, as described earlier. 
One difference between the two methods is that the rate of annihilation is likely less frequent in DMQMC than in FCIQMC because there are more density matrix elements than there are terms in the wavefunction coefficient vector.
Blunt \emph{et al.}\cite{blunt_density-matrix_2014} suggested that because of this slower rate, a higher number of walkers would be needed in DMQMC to overcome the sign problem   -- approximately the square of the size of the FCIQMC critical walker population -- and this observation was based on a Heisenberg model calculation. %

The annihilation plateau for stretched H$_6$/STO-3G is shown in \reffig{fig1}. 
This plateau occurs after the first exponential growth, when the population reaches a system-specific population of walkers, as seen in \reffig{fig1a} (for one $\beta$ loop) between $\beta = 0$ and $\beta = 5$.  
When the specific population of particles is reached, the spawning and annihilation rates are approximately equal, resulting in no population growth, i.e., the plateau. After exiting the plateau around $\beta = 15$, we observe a second exponential growth phase.
The plateau can almost always be visually identified by its distinctive appearance, although in practice, we have also automated this initial measurement (see \refsec{methods}). 

When inspecting \reffig{fig1b}, the instantaneous energy estimate begins the simulation in reasonable agreement with the finite temperature full configuration interaction (ft-FCI) energy, but quickly thereafter the energy fluctuates considerably. 
After the simulation exits the plateau region, we see a return to agreement between the DMQMC energy and the ft-FCI energy. 

We can also compare the plateau heights for stretched H$_6$/STO-3G (200 determinants) in DMQMC and FCIQMC. Here, the plateau height is measured at $2.927(5) \times 10^4$ particles for DMQMC. This system in FCIQMC has a smaller plateau height, at only $2.2(1) \times 10^2$ particles.
This is consistent with the description of Blunt \emph{et al.},\cite{blunt_density-matrix_2014} where the authors commented that the DMQMC plateau height is approximately the square of the FCIQMC plateau height.
The plateau occurs between $\beta=5$ and $\beta=15$ in this simulation and this temperature range is something that we cannot easily control as an independent variable.
Thus, while the critical temperature is something we could measure we generally neglect it for this study. 

In summary, the annihilation plateau in DMQMC for \emph{ab initio} systems follows previous observations based on model Hamiltonians\cite{blunt_density-matrix_2014} and the FCIQMC annihilation plateau. 

\subsection{Connection between the plateau in symmetric DMQMC  and the unphysical Hamiltonian and annihilation rate}\label{unphysical-ham} 

\begin{figure*} %
\begin{center}
   \subfigure[\mbox{}]{\includegraphics[width=0.45\textwidth,height=\textheight,keepaspectratio]{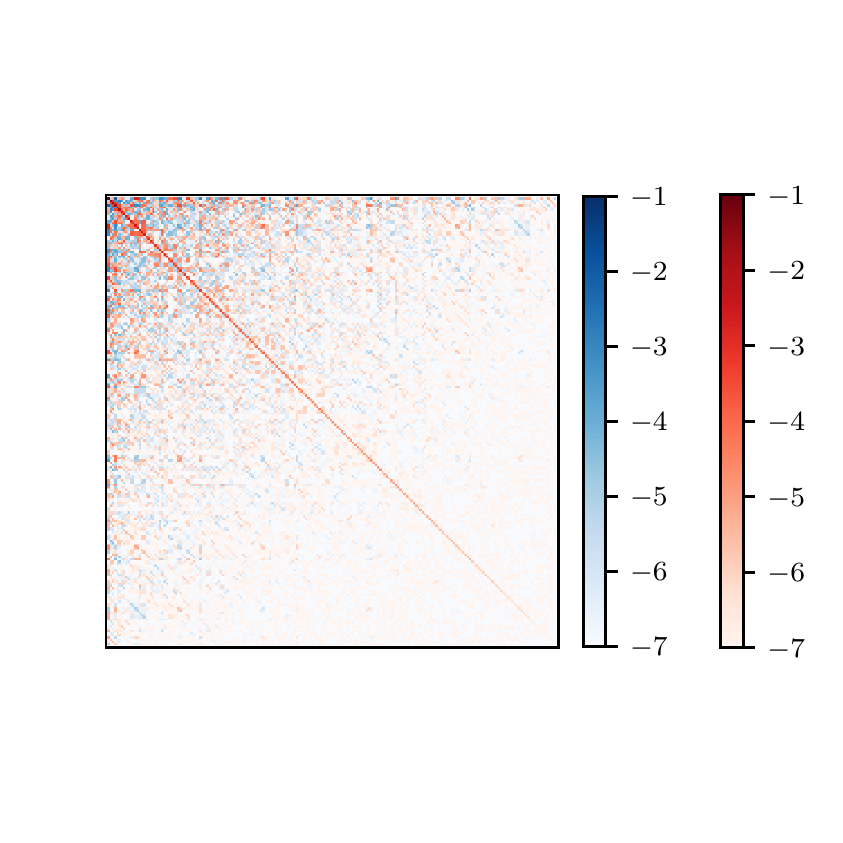}
   \label{fig2a}}
   \subfigure[\mbox{}]{\includegraphics[width=0.45\textwidth,height=\textheight,keepaspectratio]{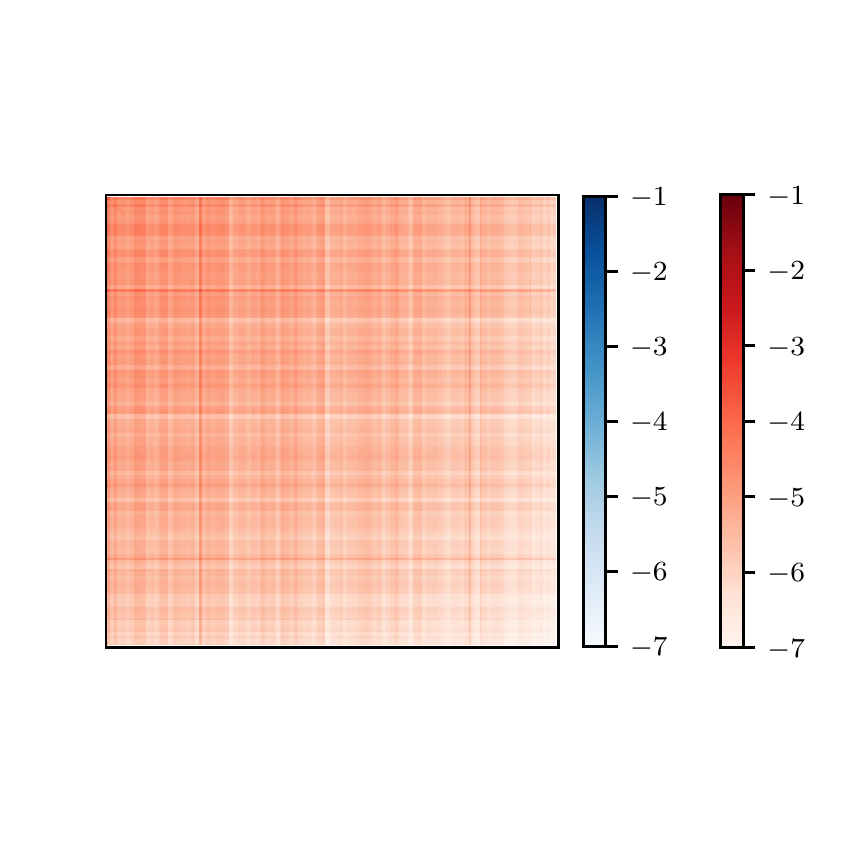}
    \label{fig2b}}
    
    \subfigure[\mbox{}]{\includegraphics[width=0.45\textwidth,height=\textheight,keepaspectratio]{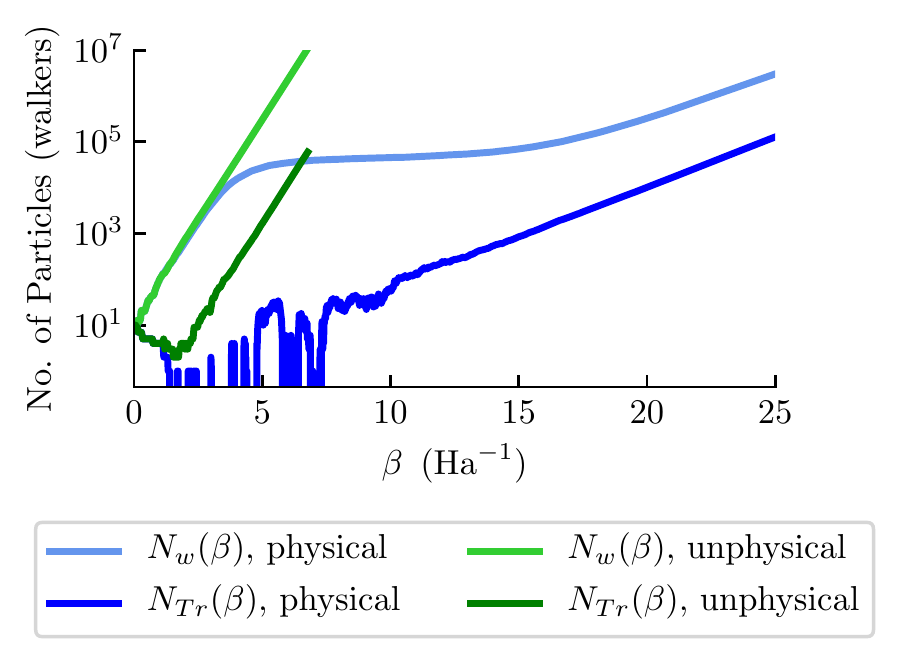}
    \label{fig2c}}
    \subfigure[\mbox{}]{\includegraphics[width=0.45\textwidth,height=\textheight,keepaspectratio]{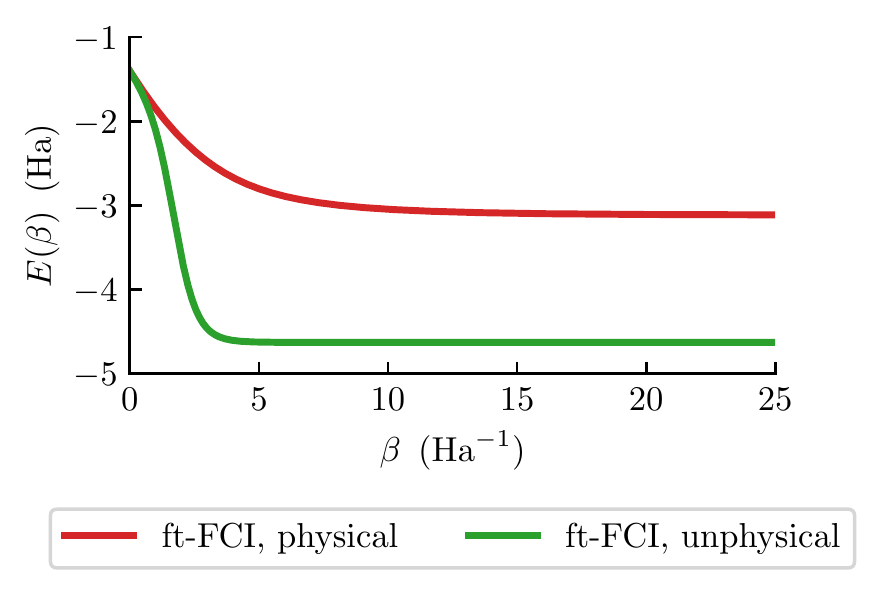}
    \label{fig2d}}
     \caption{For the stretched H$_6$/STO-3G system we show the deterministic density matrices for $\beta=3$ expressed as heatmaps for (a) the physical Hamiltonian and (b) the unphysical Hamiltonian. In (a) and (b), blue corresponds to negatively signed elements, and red corresponds to positively signed elements. The darker the color, the larger the weight of the element (based on a log scale, \emph{e.g.}, -7 represents elements of $10^{-7}$). In (c), for the same system, the population on the diagonal ($N_{Tr}$) and total walker population ($N_w$) are shown for the physical (blue) and unphysical (green) Hamiltonians from a single $\beta$ loop simulation in DMQMC. In (d), the exact temperature dependent diagonalizations (ft-FCI) of the physical (red) and unphysical (green) Hamiltonians are shown.
     In this figure, DMQMC is symmetrically propagated. 
     }
     \label{fig2}
     \end{center}
\end{figure*} %

The sign problem arises in DMQMC because spawning events 
are affected by the sign of the Hamiltonian, $H_{ik}$, connecting two density matrix elements.
In general, the sign of the matrix element $H_{ik}$ ($i\neq k$) can be positive or negative.
One way to think about how this arises is that the Slater--Condon rules are applied by bringing the occupied orbitals in determinant $i$ and $k$ into maximum coincidence by permuting the electron indices with each permutation causing a change in sign.
It follows that $\rho_{kj}$ can also have any sign. 
A sufficient number of walkers must be present to allow for the efficient cancellation of signed spawning events arriving at $\rho_{kj}$ to resolve the sign of $\rho_{kj}$.

Spencer \emph{et al.}\cite{spencer_sign_2012} proposed that the sign problem in FCIQMC was: (1) due to an unphysical Hamiltonian ($\tilde{\hat{H}}$) whose off-diagonal matrix elements have been wholly negated while leaving the magnitude unchanged, \emph{i.e.},  $\tilde{H}_{ik}=\delta_{ik}{H}_{ik}-(1-\delta_{ik})|{H}_{ik}|$, where $\delta_{ik}$ is the Kronecker delta, 
and (2) tending to be as severe as the energy of the dominant eigenvalue of the unphysical Hamiltonian.
The authors found that these can be summarized by the following equation for the critical walker population, $N_c$:
\begin{equation}
N_c\approx\frac{V_\mathrm{max}}{\kappa}
\label{eq:plateau}
\end{equation}
Here, $\kappa$ is the annihilation rate constant and $V_\mathrm{max}$ is the energy of the highest energy eigenstate of $V=-\tilde{\hat{H}}$ accounting for the shift correlation energy and the HF energy (i.e. $V_\mathrm{max}=V_0+S+E_\mathrm{HF}$).
The approximation in the equation refers to this being valid in the limit of a small population and to first order in $V_\mathrm{max}$. 
It will also be helpful to define a variable $T_\mathrm{max}=T_0+S+E_\mathrm{HF}$. This is the highest energy eigenstate of $T=-{\hat{H}}$ shifted by the same amount as $V_\mathrm{max}$.
Below, we test the same observations for DMQMC using the stretched H$_6$/STO-3G system.

It is first useful to identify the sign structure of density matrix for both the physical and unphysical Hamiltonians. These are shown in \reffig{fig2a} and \reffig{fig2b}, respectively for $\beta =3$.  
It can be seen in this figure that these matrices differ in both the signs of their elements as well as the distribution of the occupied elements. In the physical Hamiltonian, there is a mixture of both positively and negatively signed elements, distributed densely across the entire matrix. The combination of the heavily signed and densely packed elements explains why this inverse temperature is difficult to sample. In contrast, we see in the unphysical Hamiltonian matrix that only positively signed elements exist, and is more evenly distributed compared to the physical Hamiltonian matrix.
Now, in the DMQMC simulations of both Hamiltonians, different dynamics are seen.
For the physical Hamiltonian, DMQMC exhibits a characteristic plateau shape as the total population growth rises exponentially, pauses, and then resumes (\reffig{fig2c}). 
Only when the population growth resumes does the growth of walkers on the diagonal of the density matrix start in earnest. 
In the dynamics of the simulation, we see that the walkers on the diagonal tend to spawn and then die, depleting the diagonal population. 
It is only when enough of a population exists on the off-diagonal part of the density matrix and the sign structure has been established that the diagonal population can be sustained. 
By contrast, for the unphysical Hamiltonian, DMQMC exhibits largely uninterrupted growth in both the total population and the population of walkers on the diagonal.
In this case the sign problem has occurred because $\tilde{\hat{H}}\neq \hat{H}$; this condition is necessary but not sufficient.
However, there are examples where it is sufficient to have a similarity transformation which maps the two matrices onto each other. 
Such is the case in the bipartite Heisenberg model, for example.\cite{spencer_sign_2012}

To show that the sign problem is also related to the dominant eigenvector of the unphysical Hamiltonian, ft-FCI results are shown in \reffig{fig2d}. 
We can see here that in general, the energies obtained from the two Hamiltonians are different, where the unphysical Hamiltonian energy is lower than that of the physical Hamiltonian energy. 
The one exception we see is at $\beta=0$, when the two solutions are degenerate owing to the trace being the same between the physical and unphysical Hamiltonian. 
In the low $\beta$ regime is exactly where the dynamics appear to be the most similar in terms of population growth (\reffig{fig2c}) which is consistent with the idea that the dominant eigenvalue of the unphysical Hamiltonian causes a change in the population dynamics.

\begin{figure}
\includegraphics[width=0.45\textwidth,height=\textheight,keepaspectratio]{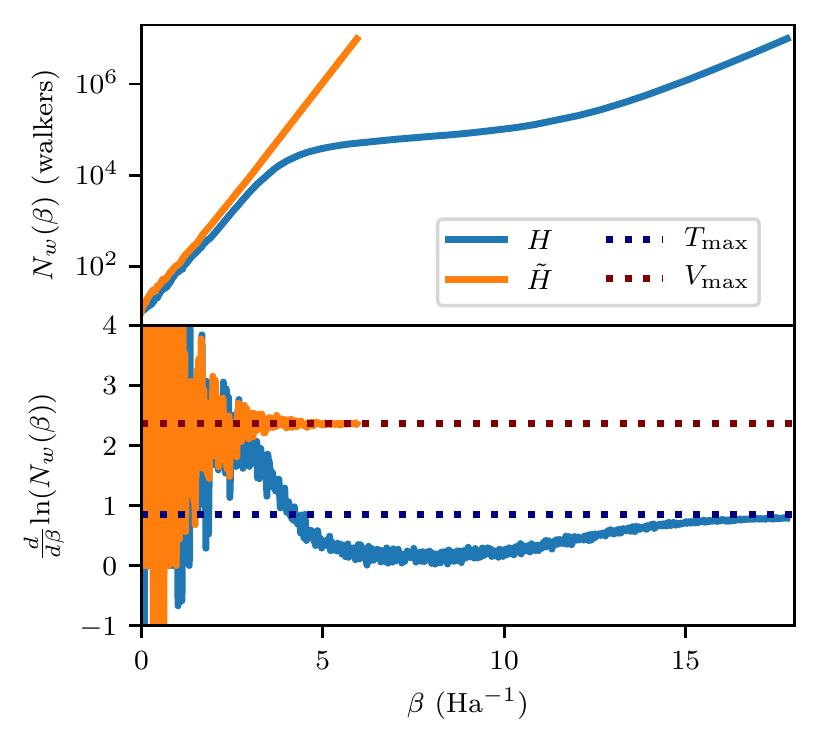}
\caption{The population growth rates for DMQMC for both $\tilde{\hat{H}}$ and $\hat{H}$ for the stretched H$_6$ system with a shift of S = 0.686. 
A single $\beta$ loop was used. $T_{\mathrm{max}}$ and $V_{\mathrm{max}}$ were found through exact diagonalization of the respective Hamiltonian matrix. In this figure, DMQMC is symmetrically propagated. }
\label{growth-rates}
\end{figure}

To analyze this further, we can compare the population growth rates when using $\hat{H}$ and $\tilde{\hat{H}}$. 
Assuming a growth rate of $N_w \sim e^{k\beta}$, we can find the instantaneous rate constant for growth from $\frac{d}{d \beta} \mathrm{ln}(N_w)$. 
This is shown in \reffig{growth-rates}. 
The growth rate for the $\tilde{\hat{H}}$ propagator oscillates around $ e^{V_{\mathrm{max}}}\beta$ for the whole of the simulation. By contrast, the growth rate for $\hat{H}$ in the pre-plateau region starts at $ e^{V_{\mathrm{max}}}\beta$, while post-plateau the growth rate tends towards $ e^{T_{\mathrm{max}}}\beta$ at large $\beta$. 
This lends further evidence to the relationship between the pre-plateau dynamics and $\tilde{\hat{H}}$.

\begin{figure} 
\begin{center}
    \includegraphics[width=0.5\textwidth,height=\textheight,keepaspectratio]{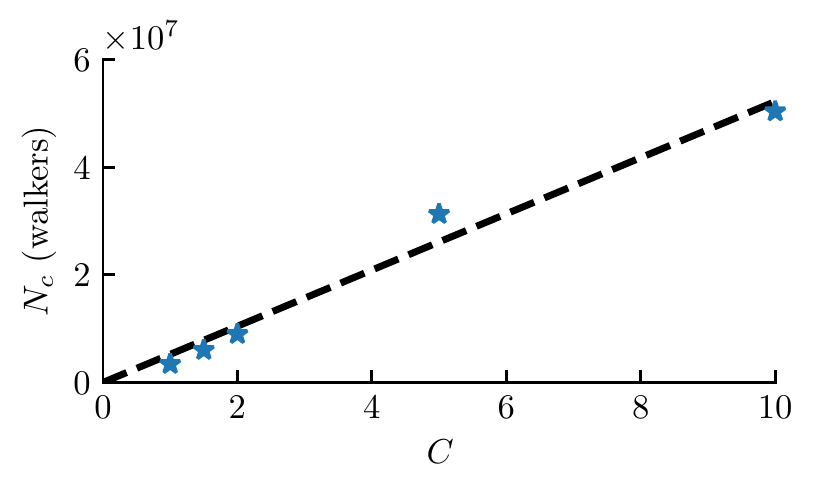}
    \caption{ The plateau height in DMQMC ($N_c$) for equilibrium H$_8$ as a function of the scaling factor $C$, described in the main text (\refeq{defineC}). Here,  the plateau heights shown are averages from four $\beta$ loops, and the values of $C$ used are $1, 1.5, 2, 5, 10$. The $y = mx +b $ fit  is $N_c =  (5.2(4) \times 10^4) \times C $, where the y-intercept  is assumed to equal zero. In this figure, DMQMC is symmetrically propagated.  }
    \label{fig3}
    \end{center}
\end{figure}

To provide further data to make the point that the dominant eigenvalue causes a change in dynamics, we scaled the off-diagonal matrix elements linearly by a factor of $C$,  starting from the true Hamiltonian, resulting in
\begin{equation}
    \tilde{H}_{ik}=\delta_{ik}{H}_{ik}+ C (1-\delta_{ik}){H}_{ik}
    \label{defineC}
\end{equation}
where $H_{ik}$ and $\delta_{ik}$ follow previous definitions, with an additional factor of a positive constant $C$. 
The plateau height for low $C$ follows a linear trend (\reffig{fig3}) which fits the form of \refeq{eq:plateau} as $V_\mathrm{max}$ is linear in $C$ for this system (assuming a constant $\kappa$).
This observation is also consistent with that of Spencer \emph{et al.}\cite{spencer_sign_2012} showing that the plateau height varies linearly with $U/t$ in the Hubbard model where $U$ is the on-site interaction strength and $t$ is the hopping integral. 
Thus the analog to $U$ in our re-scaled molecular Hamiltonian is $C$.

\begin{figure}
\begin{center}
    \includegraphics[width=0.5\textwidth,height=\textheight,keepaspectratio]{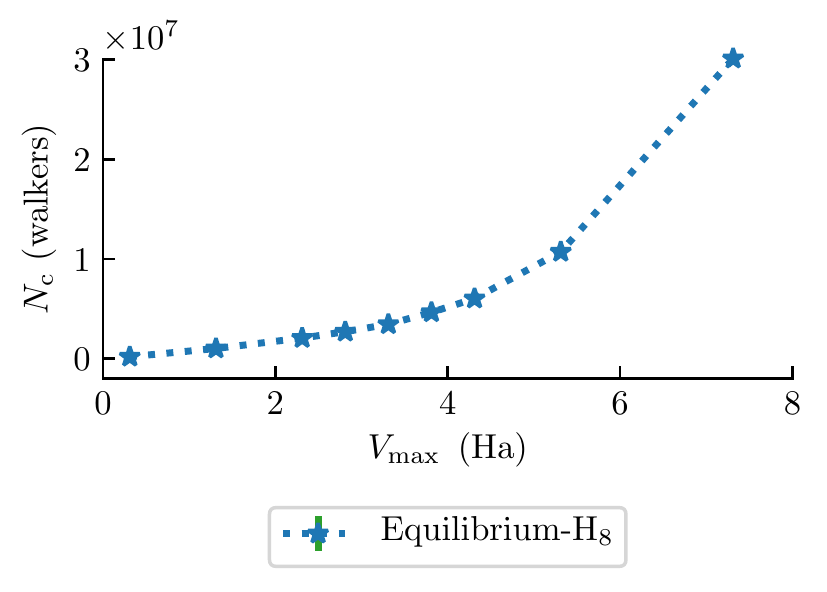}
    \caption{The critical population ($N_c$, walkers) for equilibrium H$_8$ as a function of the energy value $V_{\mathrm{max}}$, which includes the shift, $S$, as $V_{\mathrm{max}} = V_{0} +S + E_{\mathrm{HF}}$. The critical populations here were obtained from averaging over 25 $\beta$ loops. In this figure, DMQMC is symmetrically propagated. }
    \label{shift}
    \end{center}
\end{figure}

For completeness, the last component of the plateau expression given in \refeq{eq:plateau} we want to test for DMQMC is the dependence on the shift parameter, $S$. We collected data for the equilibrium H$_8$ system shown in \reffig{shift}.  
It can be seen from these data that at low $S$, the plateau height is linear in $S$ which is consistent with the form of \refeq{eq:plateau}.

In this section, we found that the sign problem and population dynamics in DMQMC can be related to similar observations made of FCIQMC.\cite{spencer_sign_2012}
In the next section we explore the relationship between the plateau heights of the two methods along with IP-DMQMC.

\subsection{How the symmetric DMQMC and IP-DMQMC plateau heights scale in relation to the FCIQMC plateau height}\label{dmqmc-plateau}

In \refsec{intro-plateaus}, we observed that the plateau height in DMQMC was approximately the square of the plateau height in FCIQMC for the stretched H$_6$ system. 
In this section, we attempt to generalize this observation to a wide range of atomic and molecular systems for both DMQMC and IP-DMQMC (which was outlined in \refsec{ip-methods}).
To achieve this, we study the range of closed-shell systems\footnote{IP-DMQMC is currently limited to treat only systems with $M_s$ = 0} that were previously considered by Booth \emph{et al.}~\cite{booth_fermion_2009} (various atoms and molecules comprised of first-row atoms) 
and supplement these with 1D hydrogen chains. 
The latter set are of interest because they are approximate analogs of the Hubbard models, which are also used for plateau studies~\cite{spencer_sign_2012,shepherd_sign_2014}. 
Thus, the total test set is comprised of: 
Ne (aug-cc-pVDZ), H$_2$O (cc-pCVDZ), HF (cc-pCVDZ), NaH (cc-pCVDZ), C$_2$ (cc-pVDZ), CH$_4$ (cc-pVDZ), N$_2$ (cc-pVDZ), stretched N$_2$ (cc-pVDZ), as well as stretched and equilibrium H$_n$ (STO-3G) for even $n$ between 4 and 16 inclusive. The Be atom is excluded from the test set as it has no measurable annihilation plateau in IP-DMQMC. 
This test set represents a variety of chemical systems including hetero- and homonuclear diatomics with single and multiple bonds. 
Our preliminary observation was that the DMQMC and IP-DMQMC plateau heights were a system-dependent fraction of the size of the space similar to FCIQMC. 
This made it difficult to establish a specific trend with system size.

\begin{figure}
\begin{center}
    \includegraphics[width=0.5\textwidth,height=\textheight,keepaspectratio]{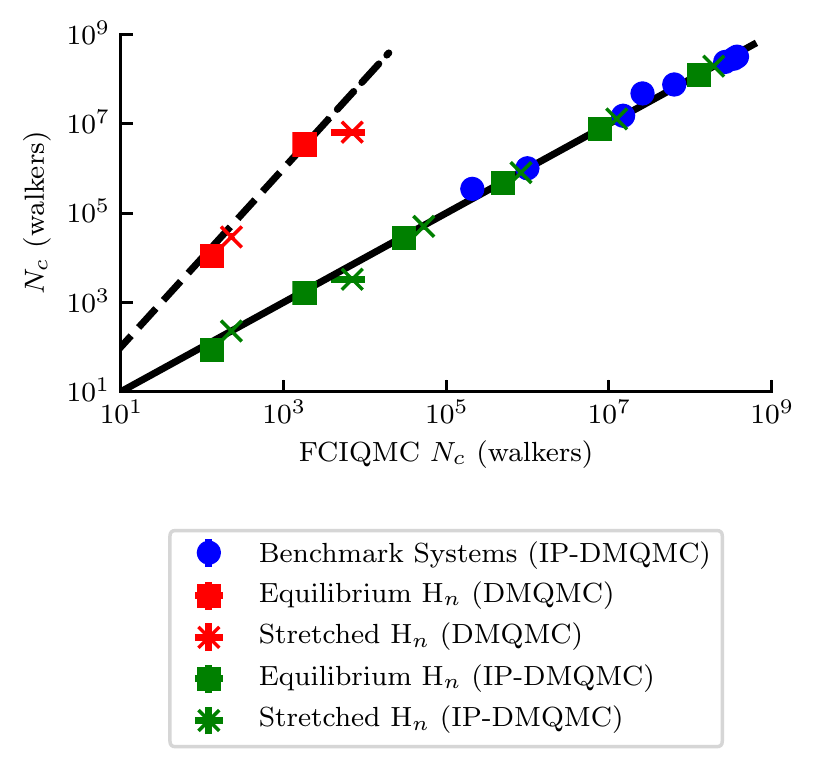}
    \caption{The plateau heights (N$_c$) for DMQMC (red) and IP-DMQMC (green and blue) simulations are shown with respect to the plateau height in FCIQMC, with both axes on a logarithmic scale for the benchmark systems from Booth \emph{et al.}\cite{booth_fermion_2009} (circle), equilibrium H$_n$ chains (even $n$ between 6 and 16 inclusive, square symbol) and stretched H$_n$ chains (even $n$ between 6 and 16 inclusive, $\times$ symbol).  IP-DMQMC simulations used a target $\beta = 25$. Straight lines are plotted for both $y=x$ (solid) and $y=x^{2}$ (dashed) to help guide the eye. The plateau heights were measured using the KDE method, and were averaged over 25 simulations. 
    The FCIQMC critical walker populations are from published data.\cite{booth_fermion_2009} Error bars are shown, and, in some cases, are smaller than the size of the marker. 
    In this figure, DMQMC is symmetrically propagated and IP-DMQMC is asymmetrically propagated. 
    }
    \label{scaling-all}
    \end{center}
\end{figure}

We anticipate that each system will have a plateau height which is a system-dependent fraction of the size of the space, similar to FCIQMC.
We therefore plot the DMQMC plateau height against the FCIQMC plateau height for the same system (Figure \ref{scaling-all}).
These values were available for equilibrium and stretched H$_6$, and for equilibrium and stretched H$_8$.  
All of the other systems in our test set had critical populations in DMQMC that were $> 5 \times 10^8$ particles (our choice of the cutoff in population in our experimental design). 
What we see in this data is that for these four systems, the DMQMC plateau height is approximately the square of the FCIQMC plateau height.  

We now turn our attention to the interaction picture variant of DMQMC (IP-DMQMC).
While this was introduced in \refsec{ip-methods}, it is instructive to provide a number of details at this point.
IP-DMQMC targets a specific $\beta$ value (here $\beta=25$ to consistently allow the plateau to be found) and initializes on an exactly known auxilliary matrix, ($\hat{f}(\tau) = e^{-(\beta -\tau)\hat{H}^0} e^{-\tau \hat{H}}$), with the weights of the auxilliary matrix replacing the random sampling of the diagonal identity matrix in DMQMC. 
IP-DMQMC also modifies the propagator such that $\hat{f}(\tau=\beta)=\rho(\beta)$ \emph{and} that the propagation is asymmetric (i.e. only happens down the rows or columns of the density matrix).

Figure \ref{scaling-all} also shows plateaus heights from IP-DMQMC.
Our calculations in \reffig{scaling-all} show that the IP-DMQMC plateau height is approximately equal to the plateau height in FCIQMC for the systems studied here. 
For example, for the stretched H$_6$ system, the IP-DMQMC plateau height is $2.2(1) \times 10^2$ particles, and the FCIQMC plateau height is also $2.2(1) \times 10^2$ particles.  
This finding is remarkable, as it shows that the critical walker population in IP-DMQMC is directly related to the same in FCIQMC. 
\begin{figure} 
\begin{center}
    \includegraphics[width=0.5\textwidth,height=\textheight,keepaspectratio]{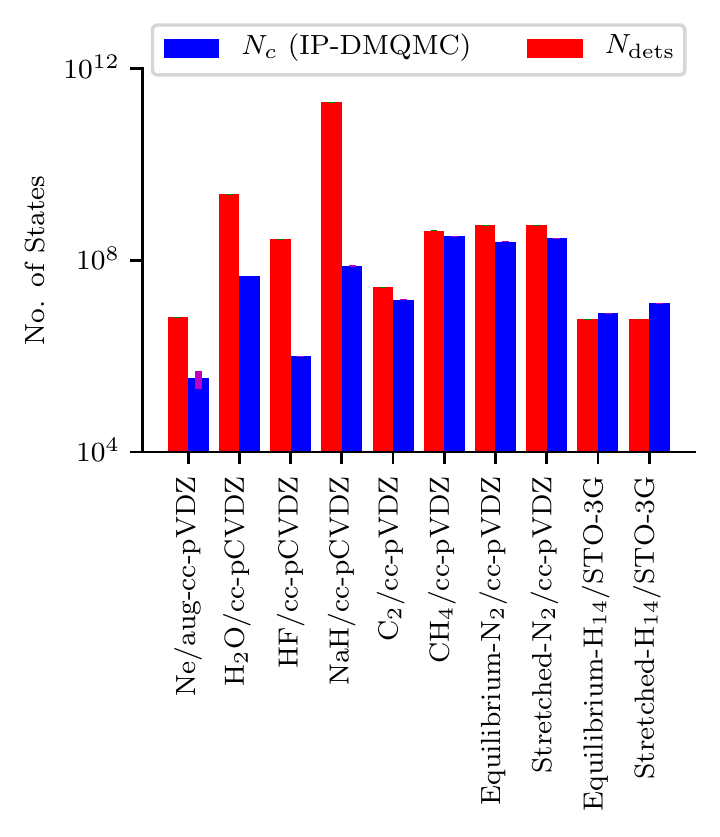}
    \caption{The critical population ($N_c$, blue) in IP-DMQMC for the atoms and molecules in our test set, compared to the estimated size of space ($N_{\mathrm{dets}}$, red). These $N_c$ were collected in the same way as \reffig{scaling-all}. In this figure, IP-DMQMC is asymmetrically propagated. 
    }
    \label{nc-ndet}
    \end{center}
\end{figure}

To further emphasize this, \reffig{nc-ndet} shows the critical walker population in IP-DMQMC plotted next to the size of the Slater determinant space in FCIQMC.
It can be seen that almost all of these systems have plateaus heights lower than the number of determinants and, therefore, lower than the square root of the number of elements in the density matrix.

\subsection{How IP-DMQMC has the same plateau height as FCIQMC}\label{ip-same-as-fciqmc}

\begin{figure} 
\begin{center}
    \includegraphics[width=0.5\textwidth,height=\textheight,keepaspectratio]{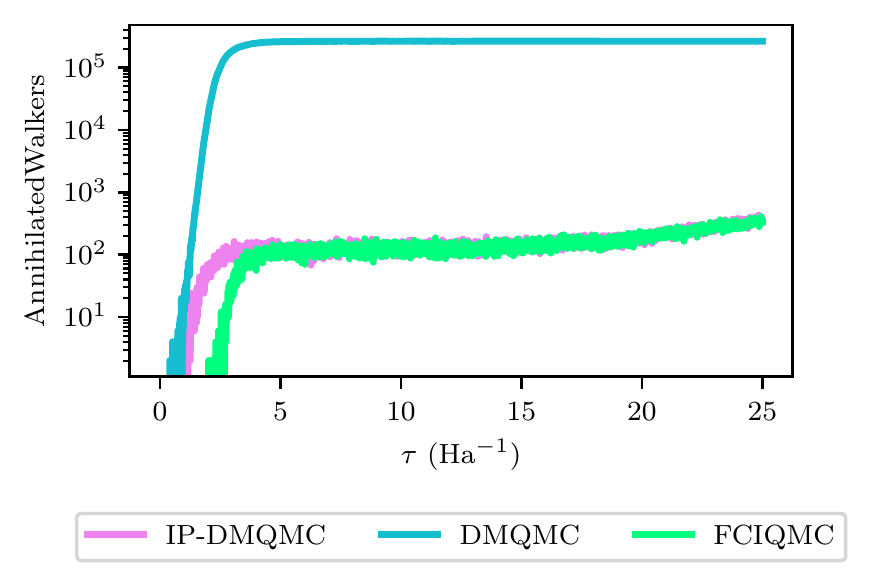}
    \caption{The number of annihilated walkers on a logarithmic scale for FCIQMC (green), IP-DMQMC (pink) and DMQMC (blue) as a function of imaginary time (iterations for FCIQMC, and $\beta$ for IP-DMQMC and DMQMC).     In this figure, DMQMC is symmetrically propagated and IP-DMQMC is asymmetrically propagated.  }
    \label{annihilation-rate}
    \end{center}
\end{figure}

In order to examine what differences in the DMQMC and IP-DMQMC methods give rise to different critical populations, we begin by analyzing \refeq{eq:plateau}. 
If we assume that $V_\mathrm{\mathrm{max}}$ is the same or approximately the same then $\kappa$ can be calculated for each method. For the simulations of stretched H$_6$, we can find $V_\mathrm{\mathrm{max}}=1.677$ Ha by diagonalization. 
Using the plateau height, %
we can then find that the $\kappa$ values for FCIQMC, DMQMC, and IP-DMQMC are $7.3 \times 10^{-3}$, $5.7 \times 10^{-5}$, and $7.3 \times 10^{-3}$ respectively. 
Here, we can see that the  IP-DMQMC rate of annihilation is the same as FCIQMC and approximately the square root of that in DMQMC \emph{i.e.} 
IP-DMQMC requires a similar rate of annihilation as in FCIQMC to resolve the sign problem.
These can be corroborated by measuring the large-$\beta$ limit growth rate of the population in \reffig{fig2c} and through measuring the annihilation rate directly from the number of walkers removed in the simulation. 
The walkers removed by annihilation are shown in \reffig{annihilation-rate}. 
The graph shows agreement with the observation above, that the annihilation rate agrees between FCIQMC and IP-DMQMC, and both are much lower than the rate in DMQMC. 

Going a step further, we can show that IP-DMQMC and FCIQMC have more similarities. Most notably, when IP-DMQMC is started from one walker, the propagator reduces to that of FCIQMC, exactly. 
To demonstrate this, we start with the IP-DMQMC propagator from \refsec{methods}, 
\begin{equation}
\frac{d\hat{f}}{d\tau} = \hat{H}^0\hat{f}- \hat{f}\hat{H}, 
\end{equation}
recalling that $\hat{H}^0$ is diagonal. 
If we assume that our one walker lands on the zeroth row, then $\hat{f}\hat{H}^0 = H_{00}f_{00}$ and will only affect the diagonal. Then, the contribution to $\Delta \hat{f}$ which is equal to $\hat{H} \hat{f}$ leads to $\hat{f}=(1+\hat{H})\hat{f}$, which is the FCIQMC propagator. 
The element $H_{00}$ refers to $\langle D_0 | \hat{H} | D_0 \rangle = E_\mathrm{HF}$.

\begin{figure} 
\begin{center}
    \includegraphics[width=0.5\textwidth,height=\textheight,keepaspectratio]{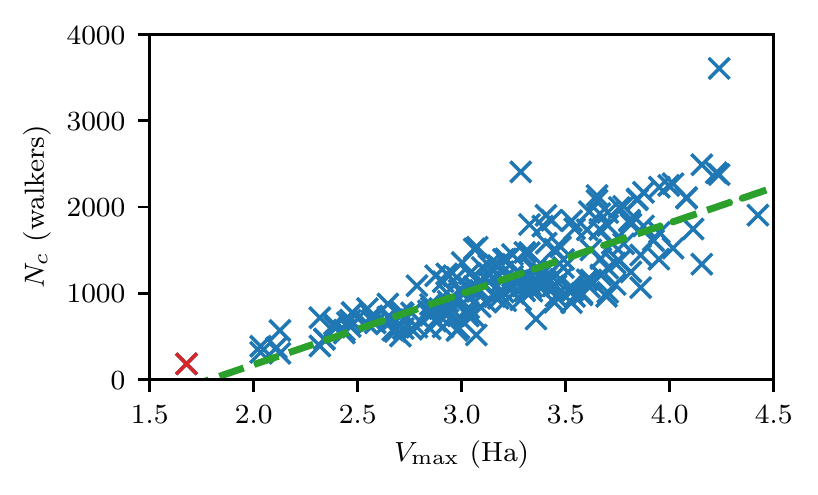}
    \caption{ The critical population of different rows in the stretched H$_6$ density matrix as a function of their $V_{\mathrm{max}}$ value. Each row has its own critical population,  and was measured from one $\beta$ loop. The zeroth row is marked with a red $\times$ symbol. Simulations for measuring the plateau height were started with one walker and had a target population of $5 \times 10^8$. $V_{\mathrm{max}}$ was calculated using an in-house analytical IP-DMQMC code, by propagating $\tilde{H}$ to $\beta = 25$ separately for each row. The energy was calculated at the beginning and end of the simulation, and $V_{\mathrm{max}}$ is equal to the difference between the final and initial energies. The linear fit (green dashed line) is given by $N_w(\beta) = 8.2(4) \times 10^2 (V_{\mathrm{max}}) - 1.4(1) \times 10^3$.  
    In this figure, IP-DMQMC is asymmetrically propagated. 
    }  
    \label{Nc-vs-Vmax}
    \end{center}
\end{figure}

In IP-DMQMC the term $H_{00} f_{00}$ modifies the Hamiltonian, subtracting the Hartree-Fock energy from the propagator, as in FCIQMC. This particular similarity between IP-DMQMC and FCIQMC is what guarantees the equivalence of the critical populations in \reffig{scaling-all}, provided that the zeroth row of $\hat{f}$ (in IP-DMQMC) is only chosen during initialization. It is reasonable to assume that when a high target $\beta$ value is used (as in our simulations shown in \reffig{scaling-all}), the zeroth row will indeed be chosen. 
Thus, \reffig{scaling-all} only represents a \emph{minimal} plateau in IP-DMQMC when the ground-state outer product is being simulated. Unless the simulation is run at very high $\beta$, we can expect that other rows will need to be simulated. 
Other rows are not encountered during an IP-DMQMC simulation started from one walker because the propagator prevents other rows from being accessed during the simulation. 
This means that we can also measure $N_c$ on a per row basis. 
When IP-DMQMC is deliberately initialized on different rows, we find that there are slight changes in the plateau as we move away from the zeroth row. 
In order to understand these changes, we note that $N_c \propto V_{\mathrm{max}}$ but that the effective $V_{\mathrm{max}}$ for a given row requires the $-\hat{H}^{(0)}\hat{f}$ term in the propagator is taken into account. 
In practice, this means that the effective $V_{\mathrm{max}}$ is raised by  $|H_{ii}|$ for row $i$.
The critical populations from different rows in IP-DMQMC are shown in \reffig{Nc-vs-Vmax} for stretched H$_6$, showing the linear relationship predicted by $N_c \propto V_{\mathrm{max}}$. 
The average critical population (taken as an average over rows) is $N_c = 1.22(3) \times 10^3$, which is slightly higher than reported in the previous section ($N_c = 2.2(1) \times 10^2$). 
The larger plateau height is due to the influence of the $-\hat{H}^{(0)}\hat{f}$ term in the IP-DMQMC propagator raising the plateau relative to $\hat{H} \hat{\rho}$ the unmodified propagator (used in DMQMC). 

\begin{figure} 
\begin{center}
    \includegraphics[width=0.5\textwidth,height=\textheight,keepaspectratio]{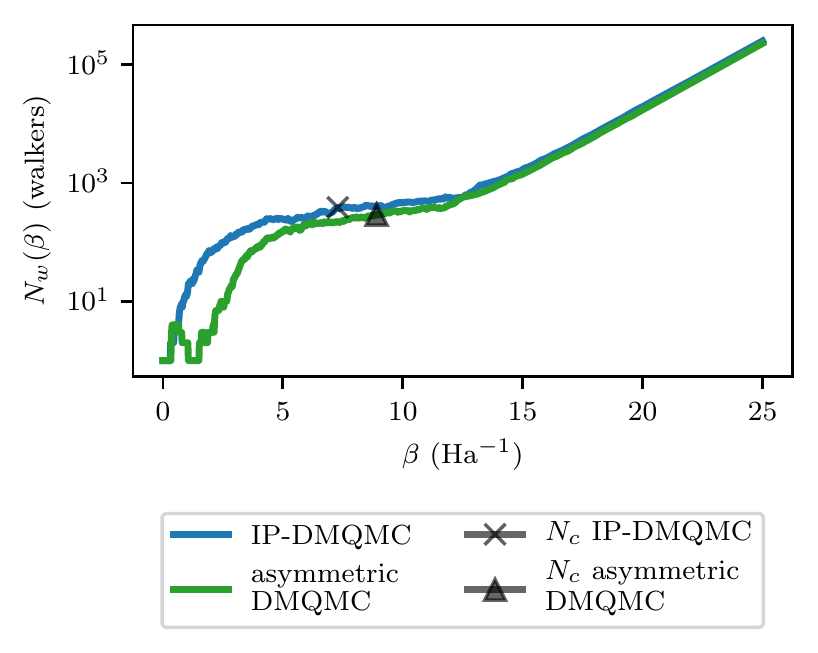}
    \caption{ The total walker populations ($N_w(\beta)$) for IP-DMQMC (blue) and asymmetric DMQMC (green) as a function of inverse temperature ($\beta$) for a random row that is not the zeroth row from \reffig{Nc-vs-Vmax}. In the asymmetric DMQMC simulation, the shift was set to $H_{ii}$ to match the IP-DMQMC methodology. The critical populations for these IP-DMQMC and asymmetric DMQMC simulations are shown as markers for the two methods (black $\times$ symbol and black triangle, respectively). Simulations were started with one walker. 
    In this figure, IP-DMQMC and DMQMC are both asymmetrically propagated. 
    } 
    \label{diagonal-shift}
    \end{center}
\end{figure}

One last question we consider in this section is the following: Does the interaction picture or the asymmetric propagation cause IP-DMQMC to have FCIQMC-like plateau heights? 
Until now, we have been running DMQMC in \emph{symmetric} mode, where rows do interact because spawning occurs along rows and columns. Recall that earlier in this section, we showed that IP-DMQMC has a lower annihilation requirement than DMQMC.
In practice, we find when DMQMC is run in asymmetric mode, it mirrors IP-DMQMC in having a lower plateau height than symmetric DMQMC. 
We actually find that the two are \emph{identical}, if the diagonal shift in asymmetric DMQMC matches what is subtracted off by $\hat{f}\hat{H}^0$ in IP-DMQMC. This is shown by visual inspection in \reffig{diagonal-shift}. 
We tested the difference by running 25 $\beta$ loops. 
We found the average plateau heights to be $N_c = 3.7(2) \times 10^2$ and $N_c = 3.44(9) \times 10^2$ for IP-DMQMC and asymmetric DMQMC respectively and the difference is not statistically significant. 
We believe, therefore, that the choice of spawning mode (symmetric versus asymmetric) gives rise to the differences we see between different methods in \reffig{scaling-all}.

In this section, we have determined the differences between DMQMC and IP-DMQMC that gives rise to different critical populations. 
A key finding was that, because the critical population measurement is started from one walker, IP-DMQMC and asymmetric DMQMC only has one row occupied throughout the whole simulation (due to the structure of the propagator). For large target $\beta$ this is likely to be the ground-state-like row making one-row IP-DMQMC equivalent to FCIQMC. 
For the non-ground-state rows, each can have a shift applied to make the plateau equivalent but, without modification, the plateau grows slightly. 
A one row asymmetric simulation on its own does not allow for a reliable thermal energy to be obtained.
To obtain an accurate thermal energy, an average over row simulations must be found e.g. by using $\beta$ loops.
The number of $\beta$ loops required to converge the energy thus plays a role in the scaling of IP-DMQMC beyond the sign problem.
Overall, then, this has the effect of allowing the distribution of memory costs across different $\beta$ loops, perhaps allowing for the convergence of systems that are too large for symmetric propagation.
However, the question remains as to how efficient this averaging is and whether there is a gain in cost relative to the DMQMC plateau. 
The subject of the stochastic error encountered when sampling the different rows of the density matrix is the subject of the next section. 

\subsection{Energy convergence with respect to number of rows and $\beta$ loops in IP-DMQMC and asymmetric DMQMC}\label{Nrows}

We now attempt to work out how many rows are required to converge an asymmetric DMQMC or IP-DMQMC calculation. 
It is, in principle, possible to converge a calculation either using row sampling (from the starting point of the simulation) or more beta loops.
We use an analytical implementation of IP-DMQMC and asymmetric DMQMC to measure the energy convergence of stretched H$_6$ with respect to $N_{\mathrm{rows}} \times N_{\beta}$ by carefully controlling the type of sampling, the number of rows and the number of $\beta$ loops. 
In the analytical code, walkers are initialized by randomly placing walkers (one at a time) on diagonal elements of the density matrix. The random distribution is uniform for asymmetric DMQMC and normalized thermal weights are used for IP-DMQMC. 
The propagation steps are handled deterministically, which removes the sign problem and allows us to isolate how many rows need to be sampled.
The error is calculated in the normal way, using analysis tools provided in the HANDE-QMC package.

The data set for asymmetric DMQMC consisted of: $N_\beta=2$, 5, 10, 20, 50, 100, 200, 500, and 1000; $N_\mathrm{rows}=1$, 2, 5, 10, 20, 50; for  $\beta=1$ to 10 in integer steps.
For IP-DMQMC, instead of fixing the number of rows the number of initialization attempts was fixed at: $N_\mathrm{attempts}=1$, 4, 40, 100, 300, 3000. 
This corresponded to approximately the same $N_\mathrm{rows}$ as DMQMC at $\beta=7$. 
The value $N_\mathrm{attempts}$ can be controlled in the original HANDE implementation through walker number. 

\begin{figure} 
\begin{center}
     \subfigure[\mbox{}]{\includegraphics[width=0.4\textwidth,height=\textheight,keepaspectratio]{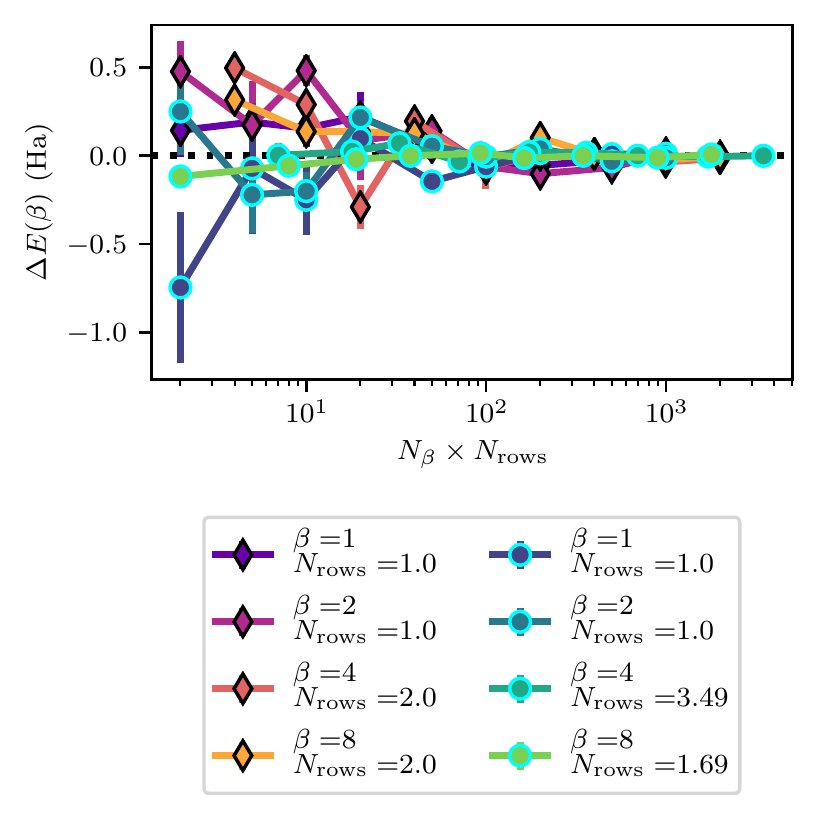}\label{deltaE}}
     \subfigure[\mbox{}]{\includegraphics[width=0.4\textwidth,height=\textheight,keepaspectratio]{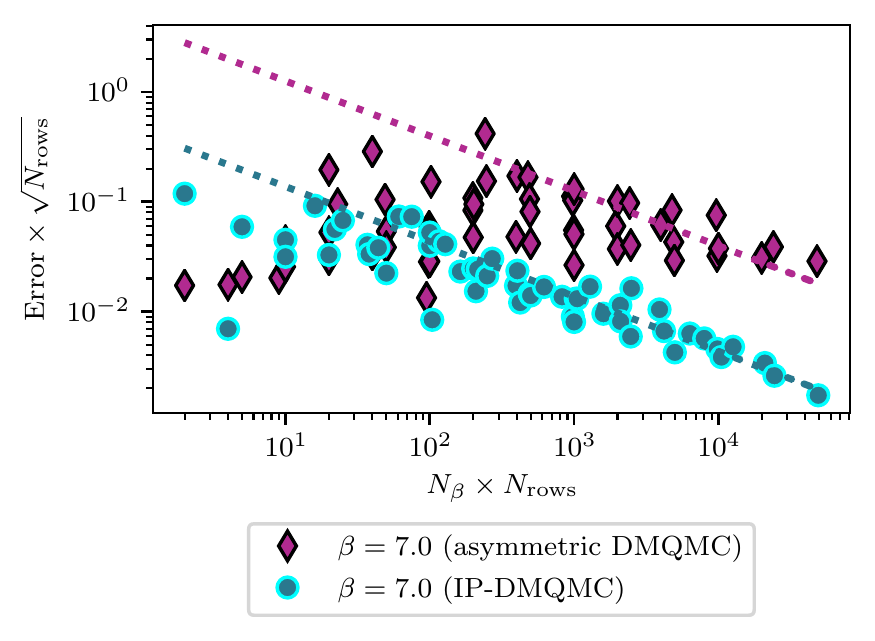}\label{err7}}
     \subfigure[\mbox{}]{\includegraphics[width=0.4\textwidth,height=\textheight,keepaspectratio]{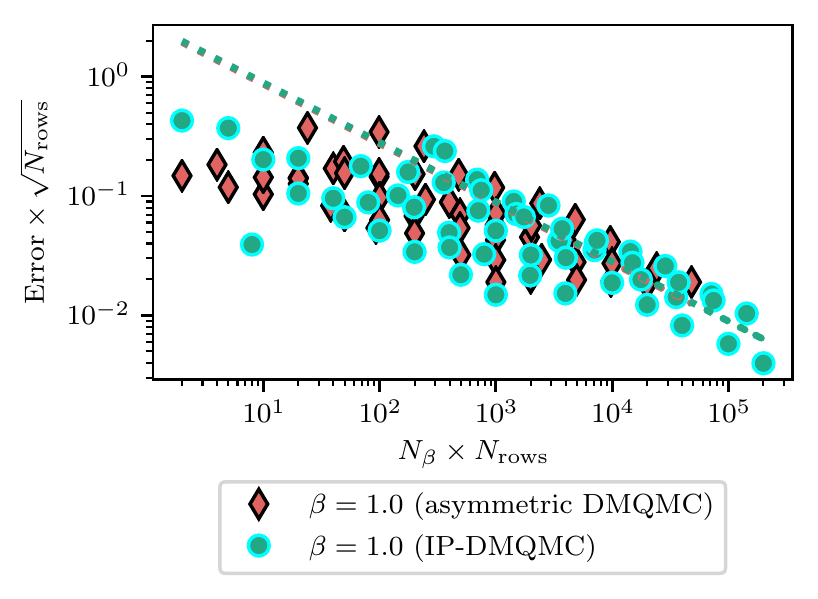}\label{err1}}
    \caption{Analytical IP-DMQMC (circles with cyan outline) and asymmetric DMQMC (diamonds with black outline) simulations of stretched stretched H$_6$.
    (a) For a minimal number of rows, convergence to the exact energy within error is possible for all $\beta$ values. (b) Stochastic error for $\beta=7.0$ shows IP-DMQMC has lower error than asymmetric DMQMC. (c) Stochastic error for $\beta=1.0$ has an almost identical error between IP-DMQMC and asymmetric DMQMC.
    Lines of best fit depict $b \times (N_{\beta} \times N_{\mathrm{rows}})^{\frac{1}{2}}$.
    In this figure, IP-DMQMC and DMQMC are both asymmetrically propagated. 
    } 
    \label{Nrows-Nbeta}
    \end{center}
\end{figure}

In general, the energy was well converged within error bars across the whole of the data set. 
This is in part because the H$_6$ system contains only 200 rows (or that there are 200 FCI determinants) which means we were oversampling in general.
However, even for $N_{\mathrm{rows}} \times N_{\beta} < 200$, we see that the energy is converged within error ($<2\sigma$) for the majority of the data set\footnote{Four exceptions with error $>2\sigma$ appeared to be randomly distributed through the data set (of $\sim500$ points)}. 
One example of this is when a minimal number of rows is sampled, which shows that the rows can be sampled independently in IP-DMQMC and asymmetric DMQMC (\reffig{deltaE}). 
This is important because it at least means the memory requirement of the plateau storage (lowered due to asymmetric propagation) can be distributed across different instances of IP-DMQMC or asymmetric DMQMC as suggested in the previous section. 

In general, we found that there was an trade-off between $N_{\mathrm{rows}}$ and $N_{\beta}$ when it came to reducing the stochastic/sampling error.
This can be seen in graphs of the stochastic error plotted against $N_{\mathrm{rows}} \times N_{\beta}$, where all of the data sets are (by visual inspection) part of the same distribution. 
This distribution generally decays according to a power-law fit of $\sqrt (N_{\mathrm{rows}} \times N_{\beta})$ in the large $N_{\mathrm{rows}}$ or $N_{\beta}$ limit. 
Examples of this are shown for two representative $\beta$ values in \reffig{err7} and \reffig{err1}. 
On these graphs, the error has been multiplied by $\sqrt(N_{\mathrm{rows}})$ to make fair comparison between IP-DMQMC and asymmetric DMQMC.
In the graphs shown we also see that $\beta=1.0$ generally has a higher error than $\beta=7.0$. 
For $\beta=7.0$, IP-DMQMC has lower stochastic error than asymmetric DMQMC by an order of magnitude while at $\beta=1.0$ their error is more comparable. 
This advantage appears to be reduced at low $N_{\mathrm{rows}} \times N_{\beta}$, which is the limit we want to be able to run our calculations in. 
It is still possible to see (in \reffig{deltaE}) that IP-DMQMC has a lower systematic error, indicating that the asymmetric DMQMC error may have an under-sampling error. 

Overall, we find that it is possible to take maximal advantage of the reduced plateau height of IP-DMQMC by running simulations on individual rows of the density matrix and averaging over $\beta$ loops. 
For higher temperatures (or for asymmetric DMQMC), the whole diagonal of the density matrix must be sampled which is likely to be costly.
However, for lower temperatures (higher $\beta$) IP-DMQMC generally converges with a smaller systematic difference to the exact result, and a smaller stochastic error. 
IP-DMQMC will exhibit reduced computational cost compared to asymmetric DMQMC due to a need to sample fewer rows (whether through walkers or $\beta$ loops).

\subsection{The initiator approach applied to IP-DMQMC}\label{initiator}

The initiator adaptation (\refsec{init-methods}) was developed to maintain population on the diagonal which is popular in FCIQMC because it removes the requirement that the simulation has to have a total walker number greater than the critical walker population (\emph{i.e.} the plateau is removed), introducing only a modest error. 
Unfortunately, the removal of the plateau means we cannot compare how i-FCIQMC and i-IP-DMQMC scale using this measure alone. 
We can, therefore, use a previous study of i-FCIQMC\cite{cleland_study_2011} where a walker population threshold measure ($N_{\mathrm{thresh}}$) of 50,000 walkers on the Hartree--Fock determinant was used as the population requirement for a converged simulation. 
This threshold is analogous to measuring the plateau height because it was shown for a variety of atoms that the energy did not vary after this threshold was reached and the simulation was converged with respect to stochastic sampling,\cite{cleland_study_2011} which is the same idea as the canonical method needing to reach a critical walker population to obtain a converged energy. 
We considered but did not attempt a ``growth witness'' measure ($G$) introduced by other authors.\cite{yang_improved_2020} 

\begin{figure} 
\begin{center}
     \subfigure[\mbox{}]{\includegraphics[width=0.4\textwidth,height=\textheight,keepaspectratio]{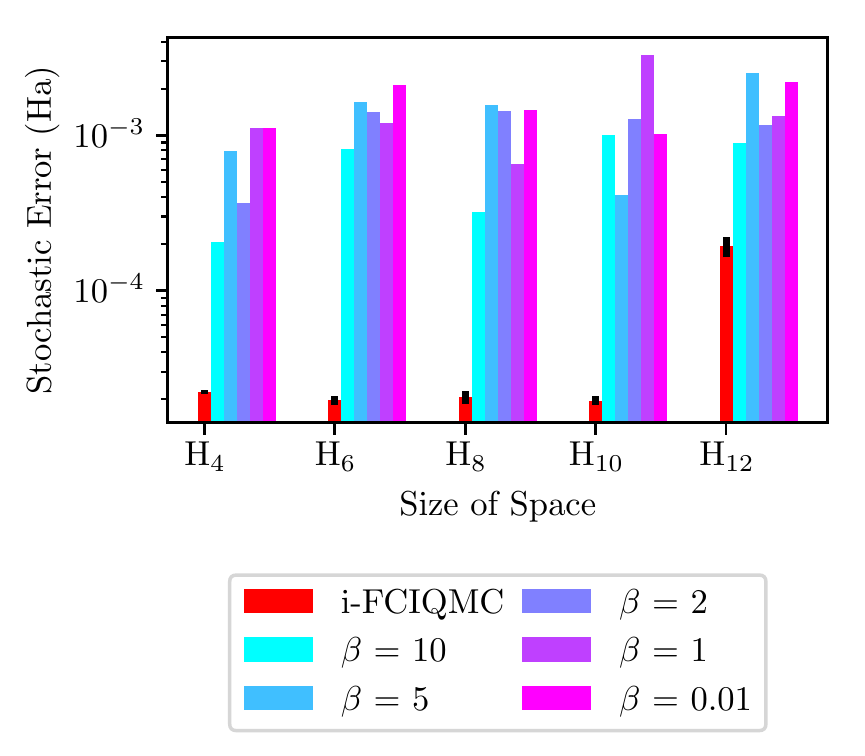}\label{fig7b}}
     \subfigure[\mbox{}]{\includegraphics[width=0.4\textwidth,height=\textheight,keepaspectratio]{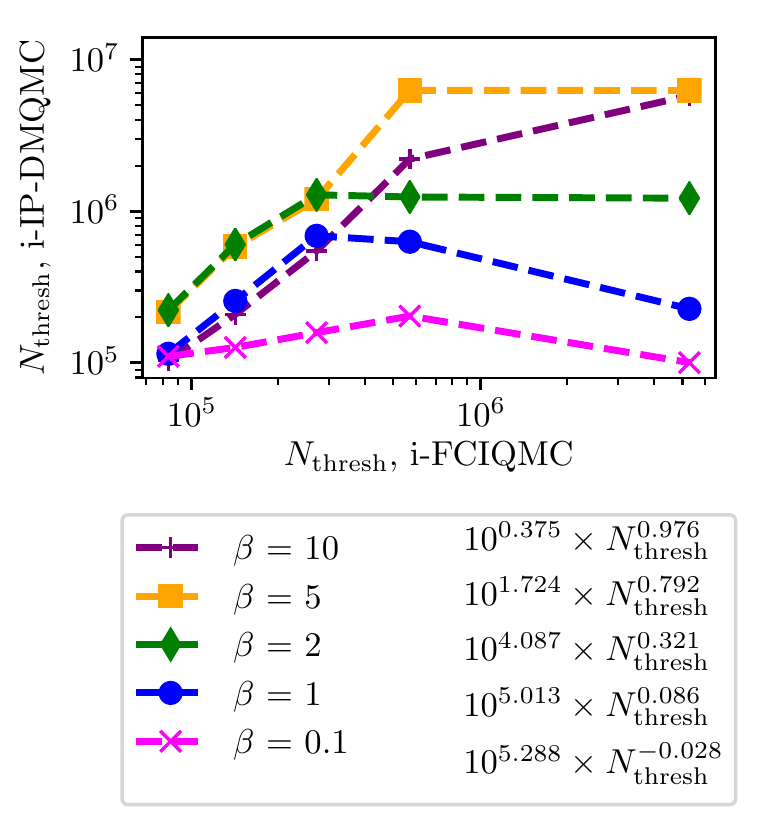}\label{fig7a}}
     \caption{
     (a) The stochastic error in i-FCIQMC and i-IP-DMQMC for linear H$_n$ chains in the STO-3G basis set, for $n = 4, 6, 8, 10, 12$. Simulations of both methods were performed at increasing target populations starting at 100 walkers and increasing to $5 \times 10^6$ walkers. For each method, a single simulation was used to determine the smallest target population ($N_{\mathrm{thresh}}$) required to reach 50,000 walkers on HF. For i-IP-DMQMC, the stochastic error shown was obtained by averaging over 5 $\beta$ loops that reached $N_{\mathrm{thresh}}$. For i-FCIQMC, the stochastic error was found for 5 different simulations (50,000 report cycles, a timestep of 0.001 and an initial population of 10 walkers), and then averaged. Error bars show one standard error. No shift damping was used in these simulations. 
     (b) The total walker population at the simulation iteration at which the population threshold was met on the diagonal of the matrix (HF determinant for i-FCIQMC) in the i-IP-DMQMC simulation ($N_\mathrm{thresh}$) as a function of the  same in the i-FCIQMC simulation ($N_\mathrm{thresh}$). 
     with initial populations of 10 particles, and with target $\beta$ values of 0.1 (magenta $\times$ symbols), 1 (blue circles), 2 (green diamonds), 5 (gold squares) and 10 (purple + symbols). The $y=10^b \times x^m$ fits are shown in the legend, on the same line as the marker corresponding to the $\beta$ value.
     In this figure, IP-DMQMC is asymmetrically propagated. 
    } 
    \label{fig7}
    \end{center}
\end{figure}

To adapt this measure for IP-DMQMC, we consider a threshold of 50,000 walkers on the trace of the density matrix which controls for systematic and stochastic errors simultaneously.
We found that this threshold provides simulations with a mostly consistent stochastic error across system sizes (\reffig{fig7b}). 
In this section, we compare i-IP-DMQMC with i-FCIQMC. 

In \reffig{fig7}, we show the total walker population at the simulation iteration at which the population threshold was met on the diagonal of the matrix (or HF determinant for i-FCIQMC), and we will refer to this walker value as $N_{\mathrm{thresh}}$ throughout this section. 
Here, we find that the i-IP-DMQMC cost in terms of walker number is the \emph{same} as i-FCIQMC for low temperatures (\emph{i.e.}, $\beta$=10), which makes sense because we are simulating the ground state when our choice of target $\beta$ is large. In the intermediate temperature range ($\beta$=2 and $\beta$=5), we find that for the smaller hydrogen chains, the cost is slightly higher in i-IP-DMQMC, but as the length of the chain is increased, i-IP-DMQMC returns to being approximately the same cost as i-FCIQMC.  At the two higher temperatures ($\beta$=1 and $\beta$=0.1), we find the cost is \emph{lower} than that of i-FCIQMC. We attribute this to the initiator adaption itself.  This variation tends to keep walkers on the diagonal of the density matrix for IP-DMQMC and at high temperatures more particles on the diagonal is closer to the physical solution, making it easier for IP-DMQMC to simulate. In general, these data show that the initiator approximation in IP-DMQMC controls the walker population in a similar manner to the initiator approximation in FCIQMC. This gives us confidence that the initiator approximation can be used in future applications. 

We note in passing that the importance sampling of DMQMC was also designed to maintain particles on the diagonal of the density matrix,\cite{blunt_density-matrix_2014} and is explored in \refapp{imp-sampling-appendix}. 
We also note that we have only looked at stochastic error here and a study of systematic initiator error is extremely important going forward.
Due to its complexities in the DMQMC method,\cite{malone_quantum_2017} %
this is left for a future study. 

\section{Conclusions}

DMQMC has been shown to be a promising method for finite temperature applications, and in this work we have confirmed that DMQMC (especially in its interaction picture variant) shows the potential to be as effective for finite temperature work as FCIQMC is for ground state simulations.  
We confirmed that the critical walker population in symmetric DMQMC scales as the square of that in FCIQMC. Additionally, we found that the critical walker population in IP-DMQMC is the same as that of FCIQMC across all $\beta$ values due to the asymmetric sampling in IP-DMQMC. We also determined that the trade-off between sampling a small amount of rows many times versus sampling all rows fewer times is approximately equal, opening an additional avenue of development for the method.
The latter is a very exciting result, as it shows that we can obtain a temperature-dependent energy at roughly the same memory and walker cost as FCIQMC, allowing us to treat systems with IP-DMQMC that cannot be treated by DMQMC. With respect to the critical walker population, this implies that IP-DMQMC has more utility compared to DMQMC, because a smaller population of particles is required to obtain the density matrix associated with the physical Hamiltonian. 
Finally, we showed that the initiator adaption with IP-DMQMC performs in a similar way than i-FCIQMC, again allowing us to expand upon the systems we can treat with this method.

As such, we now know that IP-DMQMC will be more useful than DMQMC for systems with a severe sign problem. One disadvantage of using IP-DMQMC, which we did not explore here, is that IP-DMQMC requires separate simulations to obtain energies for different inverse temperature values. Whether the computational overhead is then more expensive to obtain a full $\beta$ spectrum in IP-DMQMC compared to DMQMC is still an open question. 
We note in passing that we did not explore the connection between this observation and the Krylov projected FCIQMC,\cite{blunt_krylov-projected_2015} as we felt that this was beyond the scope of this work.

Overall, this strongly suggests that the IP-DMQMC algorithm has the same potential as FCIQMC, and gives a focus for future development. 
A natural place for future work to begin is to explore the uses (and limitations) of the initiator approach in a systematic way as well as examining ways to modify and lower the IP-DMQMC plateau height. 
For example, as it is known that basis function rotations do affect the plateau height, we are inclined to explore basis functions that are optimized for a given temperature.\cite{liu_unveiling_2020} 

\section{Acknowledgements}
Research was primarily supported by the U.S. Department of Energy, Office of Science, Office of Basic Energy Sciences Early Career Research Program (ECRP) under Award Number DE-SC0021317 (calculations and analyses by WZV and HRP). 
This work was also supported by the University of Iowa through start-up funding (research facilitation by SKR, computer time).
This research used resources of the National Energy Research
Scientific Computing Center, a DOE Office of Science User Facility supported by the Office of Science of the U.S. Department of Energy under Contract No. DE-AC02-05CH11231 (computer time).
For the purposes of providing information about input options for the calculations used, files will be deposited with Iowa Research Online (IRO) with a reference number [to be inserted at production].

\appendix
\section{The IP-DMQMC plateau at different $\beta$ values}\label{intermediate-beta-appendix}

IP-DMQMC differs from DMQMC in that a specific (target) $\beta$ must be specified and then each $\beta$ has a unique simulation.
This means that there is the potential for dependence of the IP-DMQMC plateau height on the specified $\beta$. In this section, we explore whether the IP-DMQMC plateau heights across intermediate $\beta$ values are the same as the FCIQMC plateau height. 

\begin{figure}
\begin{center}
    \includegraphics[width=0.5\textwidth,height=\textheight,keepaspectratio]{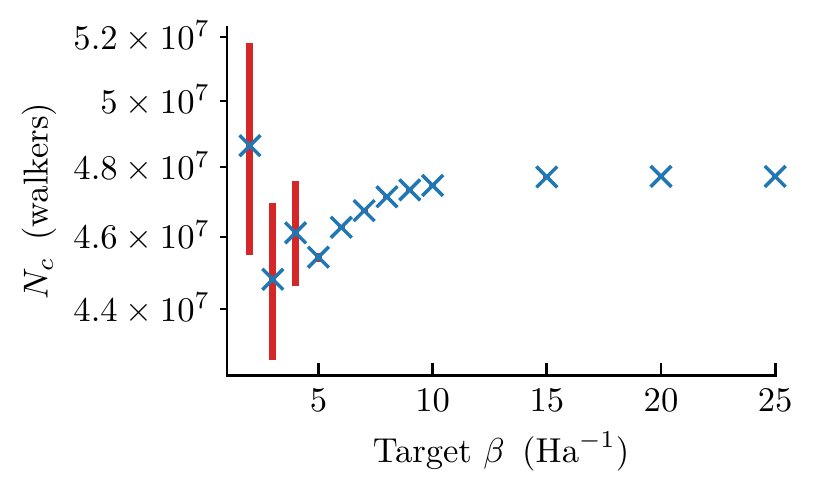} %
    \caption{The critical population ($N_c$) in IP-DMQMC as a function of the target $\beta$ value, averaged over 10 $\beta$ loops ($N_{\beta}$) for H$_2$O/cc-pCVDZ. As the $\beta$ is decreased, simulations progress through the annihilation plateau to different extents, causing an increase in the stochastic error.
    In this figure, IP-DMQMC is asymmetrically propagated. 
    } %
    \label{h2o-intermed-beta}
    \end{center}
\end{figure}

We  test this dependence using the H$_2$O system, and we present critical populations as a function of target $\beta$ in \reffig{h2o-intermed-beta}. 
As each target $\beta$ progresses through the plateau to a varying extent by the time the target is reached, to ensure a fair test we used a wall time limit of 4 hours instead. %
We found that for all $\beta$ values simulated the plateau heights are between $4 \times 10^7$ and $5 \times 10^7$, showing evidence that the IP-DMQMC critical population is not strongly $\beta$ dependent. The FCIQMC critical population for this system was found to be $4.74 \times 10^7$, and so these results confirm that the IP-DMQMC critical population is approximately the same as FCIQMC across all $\beta$ values. 

When moving to a different system, namely stretched H$_6$, we found that it was more challenging to measure a plateau height at intermediate $\beta$ values directly, as changes in input parameters would be required to make sure that a plateau even exists by the time the target $\beta$ is reached. When comparing the two systems, the walker growth is slower in the simulations of the stretched H$_6$ system compared to the growth in the simulations of H$_2$O, and we expect this may be why the plateau heights in H$_2$O can measured directly, whereas they cannot be measured directly in stretched H$_6$.  
By means of an alternative, we instead study how the (symmetric) DMQMC energy converges to the exact result with walker number and how this convergence changes a function of the target $\beta$. We do so as an alternative to changing input parameters in the IP-DMQMC simulations. 

We have, so far, established that for the stretched H$_6$ system the critical populations in symmetric DMQMC and IP-DMQMC (target $\beta=25$) are $2.92(5) \times 10^4$ and $2.2(1) \times 10^2$ particles, respectively. 

\begin{figure}
\begin{center}
\subfigure[\mbox{}]{\includegraphics[width=0.4\textwidth,height=\textheight,keepaspectratio]{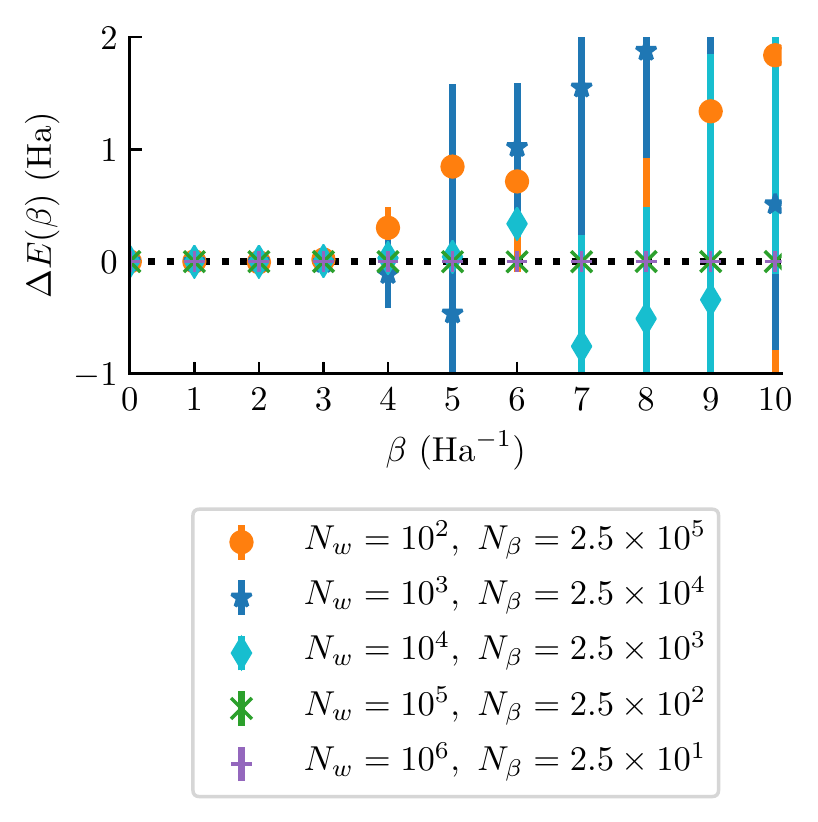}}\label{fig6a}
\subfigure[\mbox{}]{\includegraphics[width=0.4\textwidth,height=\textheight,keepaspectratio]{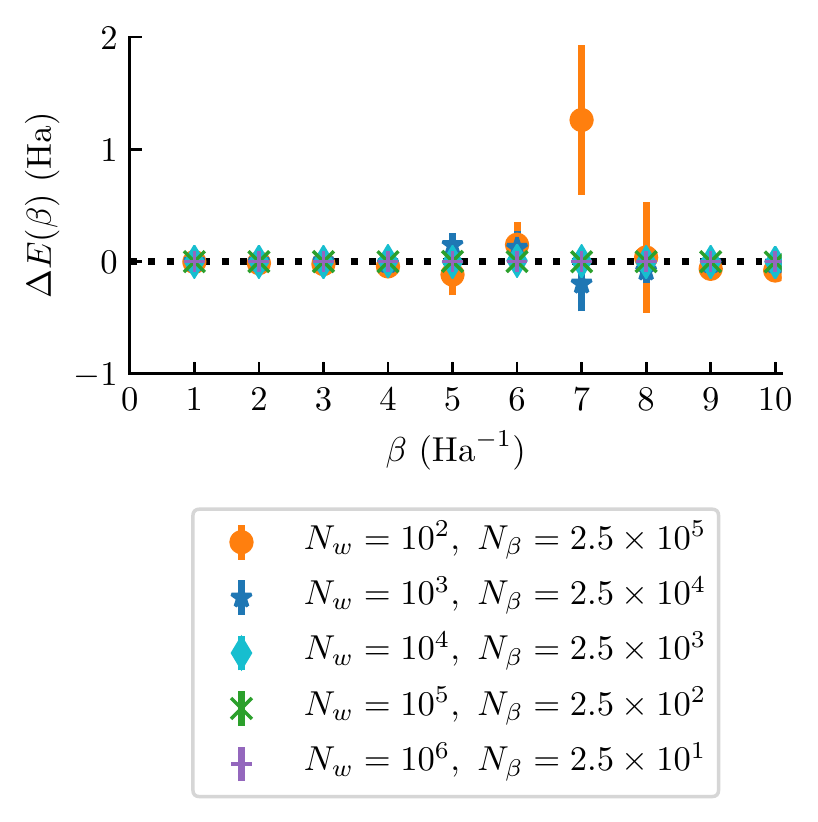}}\label{fig5a}
\caption{For the stretched H$_6$ system, the absolute energy difference from exact diagonalization (ft-FCI) is shown for (a) DMQMC and (b) IP-DMQMC. The closer to zero the energy difference is, the more converged we consider the energy. Error bars are shown but may be smaller than the size of the marker in some cases. In both figures, five populations are shown: $10^2$ (orange circle), $10^3$ (blue star), $10^4$ (teal diamond), $10^5$ (green $\times$ symbol), and $10^6$ (purple cross). For DMQMC at $N_w$ = $10^2$, the energy differences at $\beta$ = 7 and $\beta$ = 8, are outside the range of the plot, and are $-$1.814 Ha and $-$9.808 Ha, respectively.
In this figure, DMQMC is symmetrically propagated and IP-DMQMC is asymmetrically propagated. 
}
\label{fig:H6energydata}
\end{center}
\end{figure}

To test these values, simulations were performed at varying populations between $10^2$ and $10^6$ walkers with variable shift used throughout the simulation.
We note that the random initialization algorithm of IP-DMQMC means the population can vary slightly from the starting population.
For this test, the number of $\beta$ loops was reduced as the walker number was increased so the error stated is then the stochastic error for a given computational cost. 
By visual inspection of the energy differences to ft-FCI (\reffig{fig:H6energydata}), the energy differences rapidly fall to zero above $N_w=10^5$ walkers for DMQMC and $N_w=10^3$ walkers for IP-DMQMC.
After this, energies are well converged with relatively small error bars. 
This is consistent with the plateaus estimated at large $\beta$ referenced above. 
In the case of DMQMC below $N_w=10^5$ walkers, the noise grows with rising $\beta$, which is consistent with an exponentially falling signal-to-noise ratio which characterizes the sign problem. 
In particular, for DMQMC simulations below the plateau, the trace becomes very small and the energies become difficult to converge. 

\section{Importance Sampling}\label{imp-sampling-appendix}

\begin{figure}
\begin{center}
    \includegraphics[width=0.45\textwidth,height=\textheight,keepaspectratio]{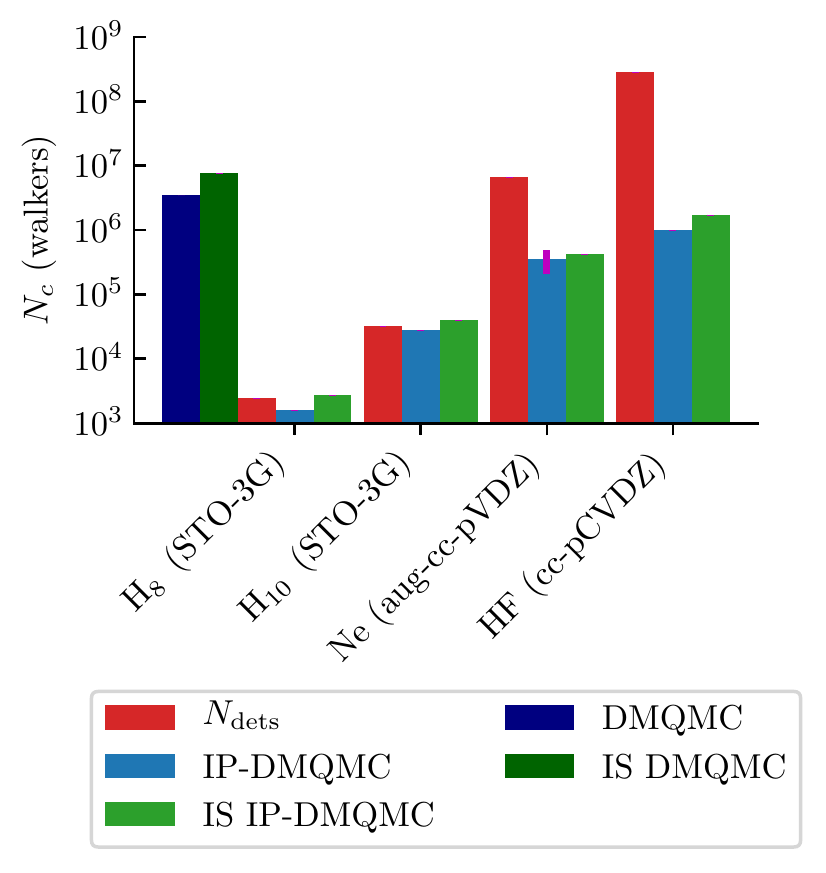}
    \caption{The plateau height ($N_c$) on a logarithmic scale for equilibrium H$_8$,  equilibrium H$_{10}$, Ne and HF in  IP-DMQMC (blue), IS IP-DMQMC (green), DMQMC (dark blue) and IS DMQMC (dark green). For comparison, the number of determinants ($N_{\mathrm{det}}$) for each system are shown (red). The critical populations in the importance sampling simulations were averaged over 25 $\beta$ loops. Importance sampling generally raises the plateau height with the exception of Ne atom where the plateau heights agree within error bars. In this figure, DMQMC is symmetrically propagated and IP-DMQMC is asymmetrically propagated.}
    \label{imp-samp-fig}
    \end{center}
\end{figure}

Importance sampling was developed along with DMQMC to prevent the escape of walkers from the trace of the simulation and improve statistical sampling; additional details can be found in Ref.~\onlinecite{blunt_density-matrix_2014}. 
The goal of importance sampling in DMQMC is to reduce the probability of particles spawning far from the diagonal\cite{blunt_density-matrix_2014}. 
Concentrating sampling on the diagonal matrix elements helps with the convergence of stochastic error as the energy expression is focused on these elements. 
Density matrix weights stored on excitations which are more than one Hamiltonian action away from the diagonal (i.e. more distant than two-particle excitations) are less likely to contribute back to the energy directly. 
The approach is to give particles on higher excitation levels larger weights, in order to avoid changes in the expectation values of the desired operator. More particles with lower weights on or near the diagonal will help to decrease stochastic error. The number of pairs of opposite signs that must be flipped in order to  reach $|D_i \rangle$ from $|D_j \rangle$ is defined as the excitation level, as first described by Ref.~\onlinecite{blunt_density-matrix_2014}. 
We find that, in general, that while the importance sampling approach does keep walkers on the diagonal of the density matrix, it also generally slightly raises the height of the annihilation plateau (\reffig{imp-samp-fig}). We note that this agrees with another study on FCIQMC in the literature.\cite{blunt_fixed_2021}
So, while it has promise in terms of converging the energy with reduced noise (due to having an increased trace population) we do not investigate it any further here.

\bibliography{js_zotero, hp_zotero, molpro, python}

\providecommand{\latin}[1]{#1}
\makeatletter
\providecommand{\doi}
  {\begingroup\let\do\@makeother\dospecials
  \catcode`\{=1 \catcode`\}=2 \doi@aux}
\providecommand{\doi@aux}[1]{\endgroup\texttt{#1}}
\makeatother
\providecommand*\mcitethebibliography{\thebibliography}
\csname @ifundefined\endcsname{endmcitethebibliography}
  {\let\endmcitethebibliography\endthebibliography}{}

 \end{document}